\documentclass[letterpaper,twocolumn,10pt]{article}
\usepackage{usenix-2020-09}

\usepackage[linesnumbered,ruled,vlined]{algorithm2e}
\usepackage[noend]{algpseudocode}

\usepackage[compact]{titlesec}
\titlespacing{\subsection}{0pt}{1ex}{0ex}
\titlespacing{\subsubsection}{0pt}{0.5ex}{0ex}
\titlespacing{\section}{0pt}{1ex}{0.4ex}

\usepackage{url}
\usepackage{cite}
\usepackage{tikz}
\usepackage{filecontents}
\usepackage{amsmath}
\usepackage{graphicx}
\usepackage{blindtext}
\usepackage{algpseudocode}
\usepackage{multirow}
\usepackage{pifont}
\usepackage{amssymb}
\usepackage{paralist}
\usepackage{xspace}
\usepackage{cleveref}
\usepackage{url}
\usepackage{soul}

\usepackage{hyperref}
\usepackage{cleveref}
\usepackage{tabularx}
\usepackage{caption}
\usepackage{subcaption}
\usepackage{xcolor}
\usepackage{multirow}
\usepackage{multicol}
\usepackage{float}
\usepackage{indentfirst}

\usepackage{tikz}
\newcommand*{\circled}[1]{\lower.7ex\hbox{\tikz\draw (0pt, 0pt)%
    circle (.5em) node {\makebox[1em][c]{\small #1}};}}

\usepackage{filecontents}
\usepackage{xspace}
\usepackage{enumitem}
\setlist[itemize]{itemsep=-0mm, leftmargin=1em}
\setlist{nosep}
\SetKwProg{Fn}{Function}{}{end}
\SetFuncSty{textsc}

\newcommand{\sysmethod}{{\em Jet\xspace}}
\newcommand{\system}{{\em Jet\xspace}}
\newcommand{\sysname}{{\em Jet Network Service\xspace}}

\newcommand{\ie}{{\em i.e.}}
\newcommand{\eg}{{\em e.g.}}
\newcommand{\etal}{{\em et al.}}

\newcommand{\todo}[1]{\textcolor{red}{TODO: #1}}

\newcommand{\qiao}[1] {\textcolor{red}{qiao: #1}}
\newcommand{\qiang}[1] {\textcolor{blue}{qiang: #1}}
\newcommand{\wyx}[1] {\textcolor{green}{wyx: #1}}

\newcommand{\para}[1]{\noindent {\bf #1}}

\begin{document}

\date{}

\title{
From RDMA to RDCA:\\ 

Toward High-Speed Last Mile of Data Center Networks \\
Using Remote Direct Cache Access
}

\author{\large 
Qiang Li$^{\diamond}$,
Qiao Xiang$^{\dagger}$, 
Derui Liu$^{\diamond}$,
Yuxin Wang$^{\dagger\diamond}$,
Haonan Qiu$^{\diamond}$,
Xiaoliang Wang$^{\ddagger}$,
Jie Zhang$^{\circ}$,\\
Ridi Wen$^{\dagger\diamond}$,
Haohao Song$^{\dagger\diamond}$,
Gexiao Tian$^{\diamond}$,
Chenyang Huang$^{\dagger}$,
Lulu Chen$^{\bigtriangleup\diamond}$,
Shaozong Liu$^{\diamond}$,\\
Yaohui Wu$^{\diamond}$,
Zhiwu Wu$^{\diamond}$,
Zicheng Luo$^{\diamond}$,
Yuchao Shao$^{\diamond}$,
Chao Han$^{\diamond}$,
Zhongjie Wu$^{\diamond}$,\\
Jianbo Dong$^{\diamond}$,
Zheng Cao$^{\diamond}$,
Jinbo Wu$^{\diamond}$,
Jiwu Shu$^{\dagger}$,
Jiesheng Wu$^{\diamond}$,
\\
$^{\diamond}$Alibaba Group, 
$^\dagger$Xiamen University,
$^\ddagger$Nanjing University,
$^\circ$Peking University,
$^\bigtriangleup$Fudan University
}

\maketitle
\begin{abstract}

We conduct a systematic measurement study on the receiver host datapath from the
	RDMA-capable NIC (RNIC) to applications in a production data center
	network (DCN).
	We find that the memory bandwidth is the bottleneck in this last
	mile of RDMA and leads to a significant drop in network throughput and a
	large increase in latency. In particular, the RNIC cannot acquire enough
	memory bandwidth due to the high contention between RNIC and CPU. 
	Consequently, in-flight packets queue up in the RNIC buffer,
	resulting in overflowed packet loss and triggering the congestion
	control mechanism even if the network bandwidth is not fully utilized.   

To tackle this problem, we propose to move the memory out of
	the receiver host datapath and reserve a small area in the last
	level cache (LLC) for the RNIC to send data to, 
 enabling remote hosts to
	directly access the receiver cache (RDCA). Our key observation is that in
	typical DCN workloads,
	the timespan data spent in memory after leaving
	the RNIC is very short (\eg, hundreds of $\mu s$ on average).
	Therefore, it is possible
	to recycle a small cache area to support RNIC
	operations at line rate (\eg, 200 Gbps). 
	We design \system{}, a
	cache-centric receiver service realizing RDCA with a
	cache-resident buffer pool, a swift cache recycle controller and a
	cache-pressure-aware escape controller. Experiments in testbed and production DCN show that
 \system{} improves throughput by up to 2.11x and P99 latency by up to 86.4\% and consumes only 12 MB LLC.

\end{abstract}

\newcommand{\tabincell}[2]{\begin{tabular}{@{}#1@{}}#2\end{tabular}}

\section{Introduction}\label{sec:intro}
Data center (DC) applications such as storage~\cite{DBLP:conf/fast/PanSZSZSPSWGCPS21,
DBLP:conf/sosp/GhemawatGL03}, high-performance  computing~\cite{DBLP:journals/pc/SpiesBMOR22,mpi}, data
analytics~\cite{DBLP:journals/cluster/BawankuleDS22,10.1145/1327452.1327492}, and machine
learning~\cite{DBLP:conf/osdi/QiaoCSNH0GX21, DBLP:conf/osdi/JiangZLYCG20} have stringent performance requirements on
high throughput and low tail latencies. Remote direct memory access (RDMA) is
an appealing solution to meet these requirements because of its low-latency,
CPU-bypassing primitives. Multiple
major DC operators have deployed RDMA in their production 
DCNs~\cite{DBLP:conf/nsdi/GaoLTXZPLWLYFZL21, DBLP:conf/sigcomm/GuoWDSYPL16,
DBLP:conf/sigcomm/MittalLDBWGVWWZ15, 
DBLP:conf/nsdi/KongZZJYGZ22}.  They design different mechanisms (\eg, congestion
control~\cite{DBLP:conf/sigcomm/ZhuEFGLLPRYZ15, kumar2020swift}, scalable
RPC~\cite{10.1145/3387514.3405897, DBLP:conf/eurosys/ChenLS19}, and new features
on programmable NIC~\cite{DBLP:conf/nsdi/Shu0CGQXCM19,
DBLP:conf/fpl/FukudaITKSAM14,
DBLP:conf/asplos/KaufmannPSAK16}) to ensure RDMA operates efficiently in
large-scale, IP-routed DCNs.  
These mechanisms focus on
smoothing RDMA operations in network fabrics (\eg, switches and RNICs), because
the conventional wisdom is that network is usually the bottleneck. 

\begin{figure*}[t]
    \centering
    \begin{subfigure}[t]{0.3\linewidth}
        \centering
		\includegraphics[width=\linewidth]{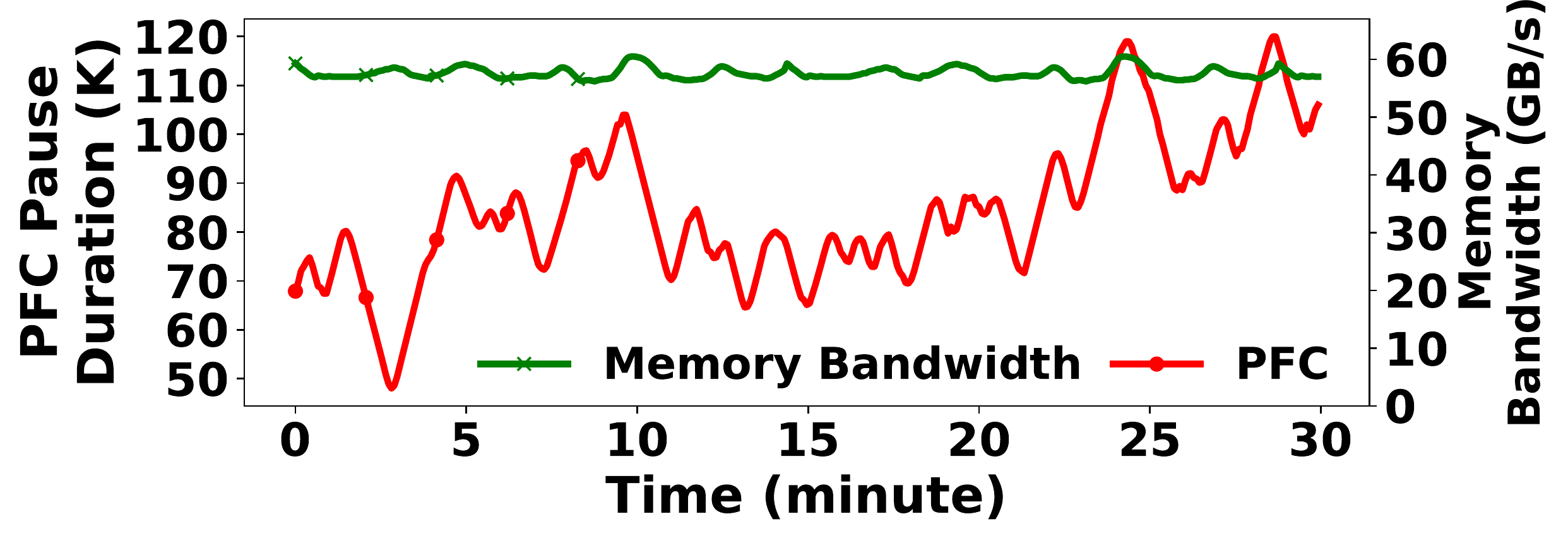}
		\caption{25 Gbps PFC-enabled sub-DCN.}
		\label{fig:moti-25lossless}
    \end{subfigure}
    \quad
    	\begin{subfigure}[t]{0.3\linewidth}
		\centering
		\includegraphics[width=\linewidth]{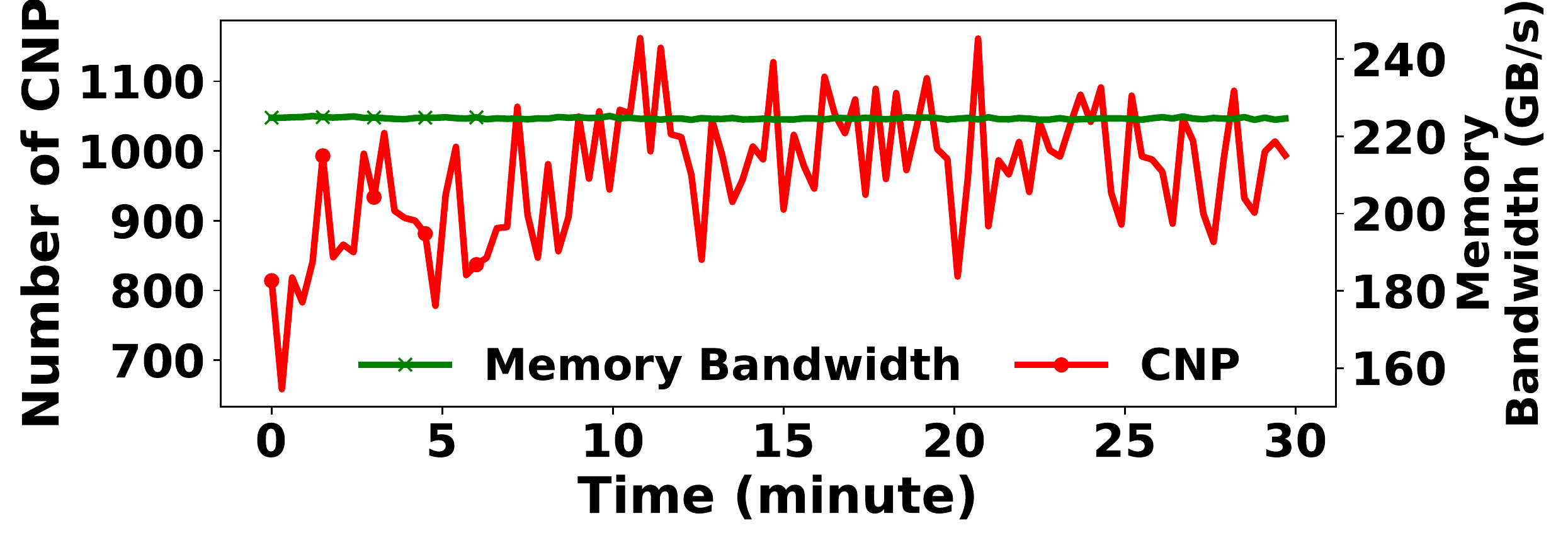}
		\caption{100 Gbps PFC-free sub-DCN.}
		\label{fig:moti-100lossy}
	\end{subfigure}
\quad
    	\begin{subfigure}[t]{0.3\linewidth}
		\centering
		\includegraphics[width=\linewidth]{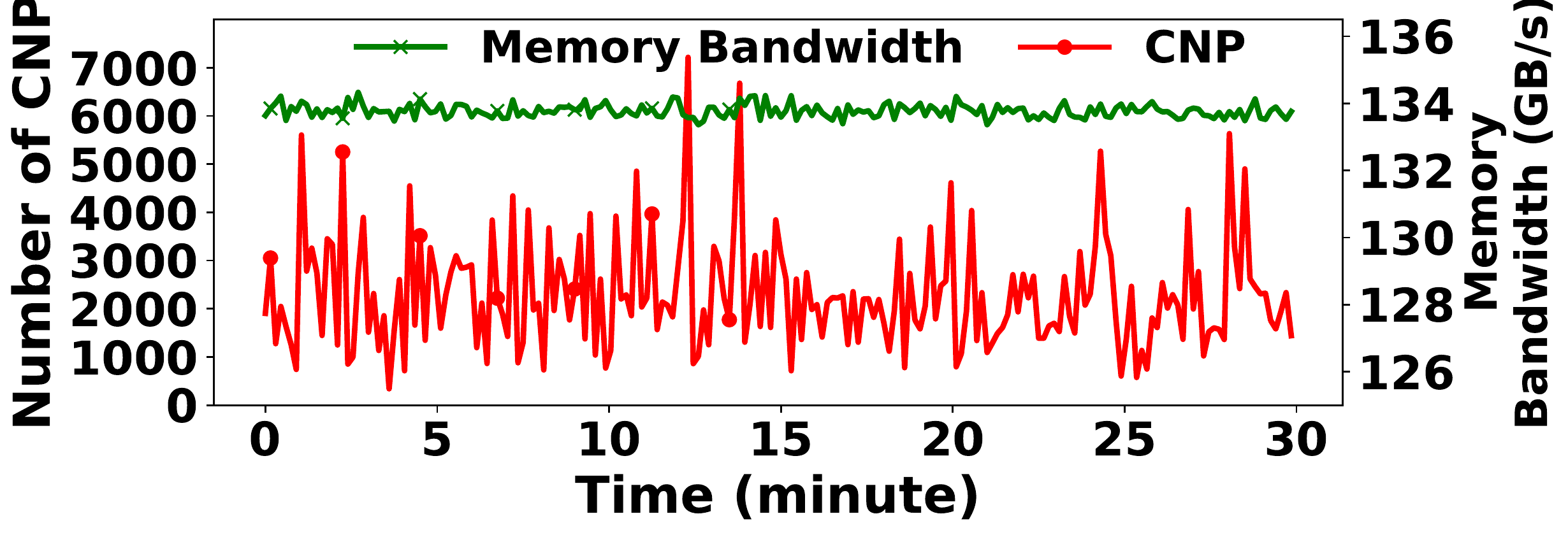}
		\caption{25 Gbps PFC-free sub-DCN.}
		\label{fig:moti-25lossy}
	\end{subfigure}    
    \caption{Memory bandwidth and PFC duration/CNP in throughput-drop incidents
	in a production DCN.}
    \label{fig:moti-online}
    \vspace{-2em}
\end{figure*}

\para{Finding in a production DCN: the memory bandwidth bottleneck limits the
performance of RDMA. (\S\ref{sec:measurement})} 
A systematic measurement study in a production DCN finds that the
memory bandwidth bottleneck in the \textit{receiver host datapath} from the RNIC
to applications limits RDMA from delivering data at a high bandwidth and a low
latency.
To be concrete, when the RNIC performs DMA operations to send received
messages to memory, it causes a high contention on memory bandwidth between RNIC and CPU
performing computations (\eg, in-memory data analytics, data replication, and
garbage collection).
Under this contention, the RNIC cannot acquire sufficient memory bandwidth to
send received packets to memory. The RNIC buffer is filled with
in-flight packets and eventually drops overflowed ones. It triggers the
 congestion control (Figure~\ref{fig:moti-online}) even if the network is not fully
utilized, leading to a large throughput drop (\eg,
\textasciitilde 15\% drop per server in the production DCN) and a considerable increase in
latency. Our recent conversations with other DC operators reveal that this
phenomenon is not an isolated finding but a universal one. One operator
recently published this issue~\cite{DBLP:conf/hotnets/0001AMMERKKRCV22}.

This memory-bandwidth-induced RDMA performance degradation can be alleviated by
optimizing the memory usage of applications and
RNIC~\cite{10.1145/3387514.3405897, recio2007remote, aifm, hydra}. However, these solutions are mostly tailored for specific applications and specific RDMA verbs.
It is also unclear whether they can scale up with the
bandwidth expansion of RDMA technologies (\eg, 100
Gbps/port in 2016~\cite{cx5_manual} and 400 Gbps/port in 2021~\cite{cx7_manual}).
Some recent studies propose redesigning RNIC by adding memory modules~\cite{netdam,DBLP:conf/micro/AlianK19}.
However, they require upgrading DCN with
specialized, non-commodity RNICs\cite{DBLP:conf/asplos/PismennyL0T22}.

\para{Proposal: move memory out of the receiver host datapath and put in the reserved cache. (\S\ref{sec:challenge})} 
Instead of continuing to optimize RNIC and applications for more efficient use
of memory bandwidth, we embrace a bold proposal: move memory out of the
receiver host datapath and reserve a small area in the cache, which has a much higher bandwidth than memory, for the RNIC to forward the
received messages to. In this way, senders do not access the receiver's
memory but its cache instead.  We coin the name remote direct
cache access (RDCA) for this design. RDCA allows the RNIC to enjoy the high bandwidth of cache. As a result, there will be no memory-bandwidth-induced
packet loss, and the RNIC can operate at line rate (\eg, 200 Gbps).

The key observation supporting the feasibility of RDCA is that in typical DCN
workloads (\eg, storage, machine learning, and HPC), the
post-RNIC timespan (\ie, the timespan data spent in memory after leaving RNIC) 
is very short (\ie, hundreds of $\mu s$ in average). As such, based on Little's law~\cite{little_law}, the average amount of
data the RNIC stores in the memory is small enough to be held in the last-level
cache (LLC) of commodity servers. For example, when the average of this
post-RNIC timespan is 200 $\mu s$, for a dual-port 100 Gbps RNIC, we only need 
5 MB for temporally storing the data leaving RNIC on average while Intel Xeon E5 provides up to 20 MB LLC.

Direct Cache Access (DCA)~\cite{DBLP:conf/isca/HuggahalliIT05} and Data
Direct I/O (DDIO)~\cite{ddio_brief} take the first steps in leveraging cache for
fast packet processing. DCA prefetches I/O
metadata to cache (\eg, descriptors and packet header)
but requires the RNIC to access
data in memory~\cite{ddio_brief,arm_caceh_stashing}. DDIO  lets RNIC access data in the LLC 
and introduces write allocate and write update operations to update
the cache. 
DDIO is deployed in Intel Xeon E5 CPU~\cite{ddio_brief}, but its benefit in high-speed networks is small. This issue,  known as the leaky
DMA problem~\cite{DBLP:conf/micro/VemmouCD22, DBLP:conf/usenix/FarshinRMK20,
DBLP:conf/nsdi/TootoonchianPLW18, DBLP:conf/isca/YuanA00KTK21}, is due to the
frequent write allocate triggered by new incoming messages, forcing
the RNIC to access data in memory. In contrast, we go beyond the passive designs
of DCA and DDIO and show that with a careful design, we can recycle a small area
of LLC to support RNIC operations at line rate.

\para{\system{}: a cache-centric service realizing RDCA
(\S\ref{sec:challenge}, \S\ref{sec:overview}, \S\ref{sec:design}).}
Although the idea makes sense on a high level, realizing RDCA has challenges. First, although the actual total space needed for temporally storing
messages leaving RNIC is small, RDMA queue pairs (QPs) over-reserve buffers,
resulting in inefficient space statistic multiplexing.  Second, reserving a
part of the LLC would cause a higher contention on the remaining LLC among other
processes. As such, it is desirable to reduce data's post-RNIC timespan so
that RDCA consumes a smaller LLC. Third, RDCA must handle
occasional jitters, such as straggler data and burst arrivals of data with a longer post-NIC
timespan.
We design \system{}, a cache-centric service tackling these challenges with three key components:

\para{A cache-resident buffer pool (\S\ref{sec:pool}).}
To address the QP buffer space challenge, we design a cache-resident buffer pool
that uses (1) SEND/RECEIVE verbs and a shared receive queue (SRQ) for receiving
small messages, and (2) the READ verb with a window-based rate control mechanism for receiving large messages, so that the total QP buffers can fit
into the reserved LLC. 

\para{A swift cache recycle controller (\S\ref{sec:recycle}).}
To consume a smaller LLC, we design a swift cache recycle controller that
reduces data's post-RNIC timespan by (1) processing data in  parallel along a pipeline and (2) optimizing processing using hardware offloading and lightweight (de)serialization.

\para{A cache-pressure-aware escape controller (\S\ref{sec:escape}).}
To handle occasional jitters, the escape controller
monitors the usage of reserved LLC and takes corresponding actions,
including (1) replacing the cache buffer used by straggler data with a new one;
(2) copying the data of slow-running applications to memory if there are too
many replacements; and (3) let the RNIC mark explicit congestion notification (ECN) in Congestion Notification Packets (CNPs) to indicate congestion if copying to memory fails or cannot release the cache pressure.

\para{Implementation and evaluation (\S\ref{sec:impl}, \S\ref{sec:eval}).}  
We evaluate \system{} extensively in testbed and production DCN using production
storage workloads.
\system{} consumes a 12 MB LLC (~20\% of
the total cache) per server and improves network throughput by up to 2.11x and P99 latency by up to 86.4\%. We also
evaluate \system{} in a DCN testbed hosting HPC workloads and find that it reduces the
average latency of collective communications in latency-sensitive HPC
applications by up to 35.1\%.

\section{Motivation}\label{sec:mot}
We first present our measurement study on the receiver host
datapath in a production DCN (\S\ref{sec:measurement}). We then
elaborate on the rationale and challenges of RDCA (\S\ref{sec:challenge}).

\subsection{Measurement in a Production DCN}\label{sec:measurement}
The production DCN consists of multiple Clos-based, IP-based sub-DCNs connected to
regional gateways.
Each has different hardware and runs RoCEv2 and TCP concurrently. Like other DCNs, the majority workload
is storage~\cite{kumar2020swift,amazon-ec2}.

\para{Intermittent receiver throughput drops due to the memory
bandwidth bottleneck at the receiver host datapath.}
We first investigate a 96-server, two-tier pod in a \textit{PFC-enabled} sub-DCN (\ie,
running DCQCN~\cite{DBLP:conf/sigcomm/ZhuEFGLLPRYZ15} and PFC~\cite{pfc}).  Each server has a dual-port 25
Gbps ConnectX-4 LX RNIC.  During its operation, we observe intermittent,
long-lasting throughput drop (\ie, up to \textasciitilde15\% server throughput
drop over 30 minutes). We analyze the logs and find that the only difference before and
after the throughput drop is the occupied memory bandwidth at the
server. 

We pick one such incident, during which the server's workload has an
approximately 1:1 read/write ratio.  Figure~\ref{fig:moti-25lossless} shows that
it consumes a close-to-capacity \textasciitilde60 GB/s memory bandwidth during the
incident.  The red line in Figure~\ref{fig:moti-25lossless} plots the PFC pause
duration of the server during the incident. We observe that the average PFC
pause duration is around 80K, with several severe bursts exceeding 100K. Such
high PFC pause duration causes the watchdog to frequently trigger PFC switch-off,
leading to the server throughput drop. 

\begin{figure}[t]
\centering
\begin{subfigure}[t]{0.45\linewidth}
		\centering
		\includegraphics[width=\linewidth]{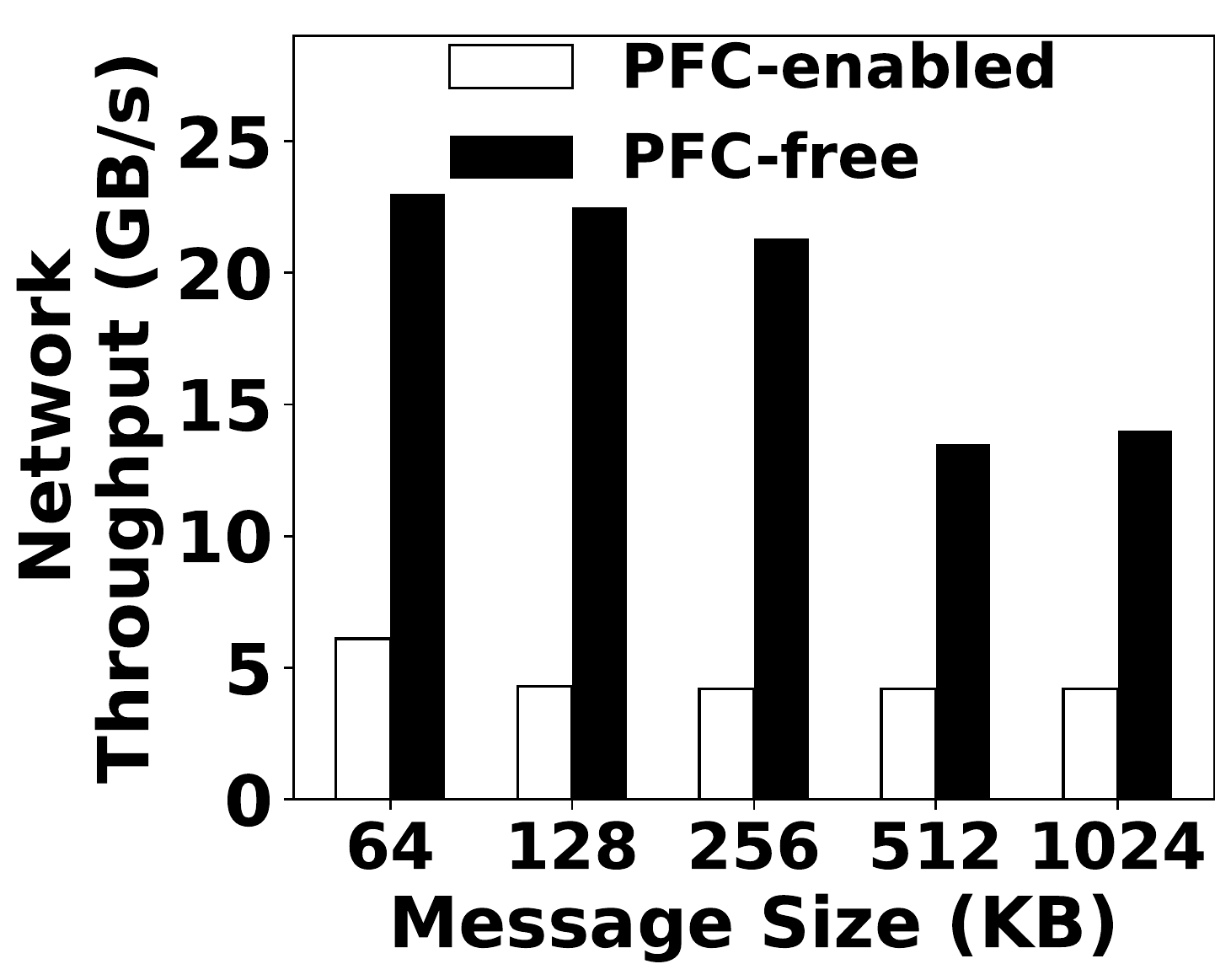}
		\caption{Receiver throughput.}
		\label{fig:moti-bw}
	\end{subfigure}    
\quad 
\begin{subfigure}[t]{0.45\linewidth}
		\centering
		\includegraphics[width=\linewidth]{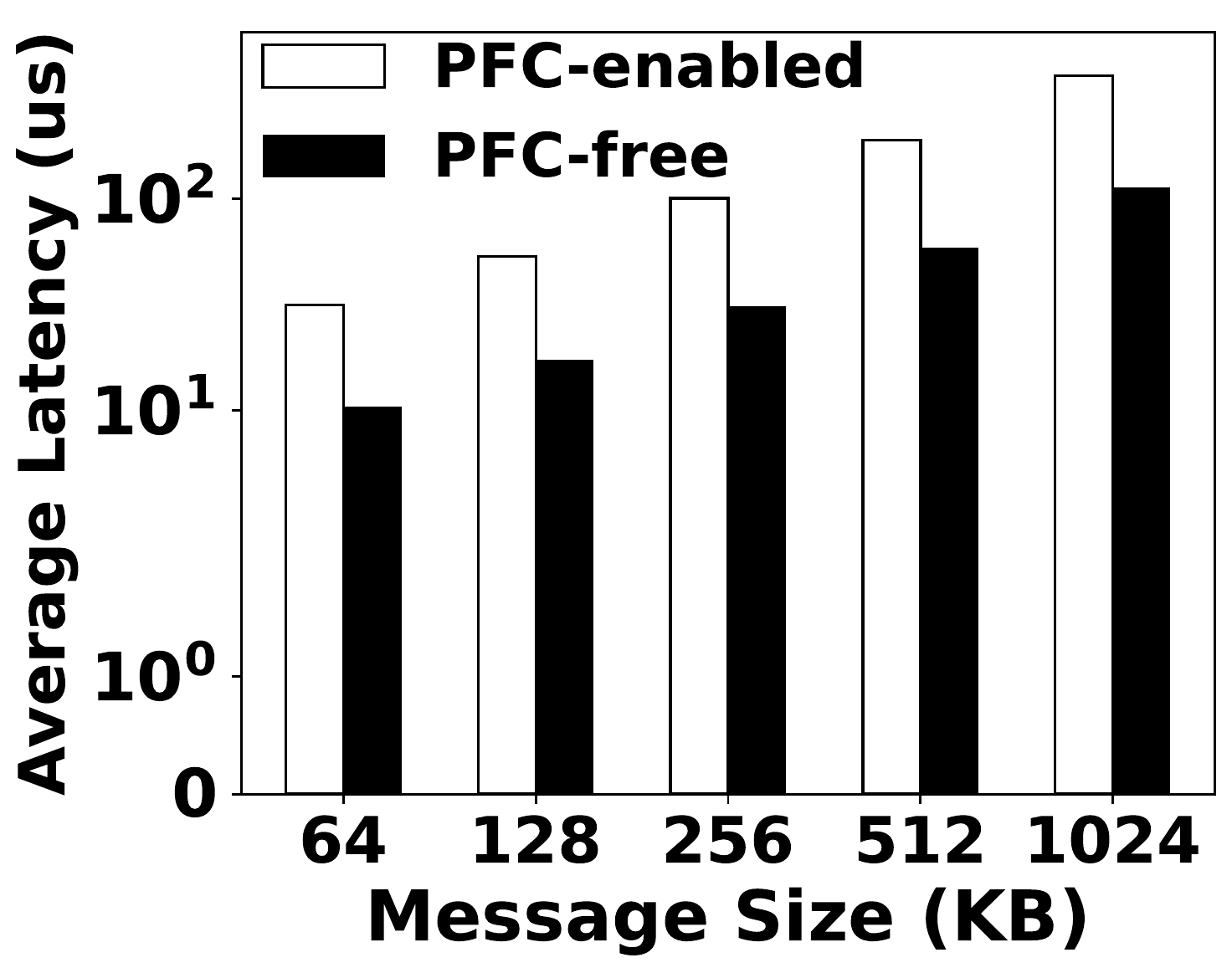}
		\caption{Network average latency.}
		\label{fig:moti-avg-lat}
	\end{subfigure}    
 \quad
	\begin{subfigure}[t]{0.45\linewidth}
		\centering
		\includegraphics[width=\linewidth]{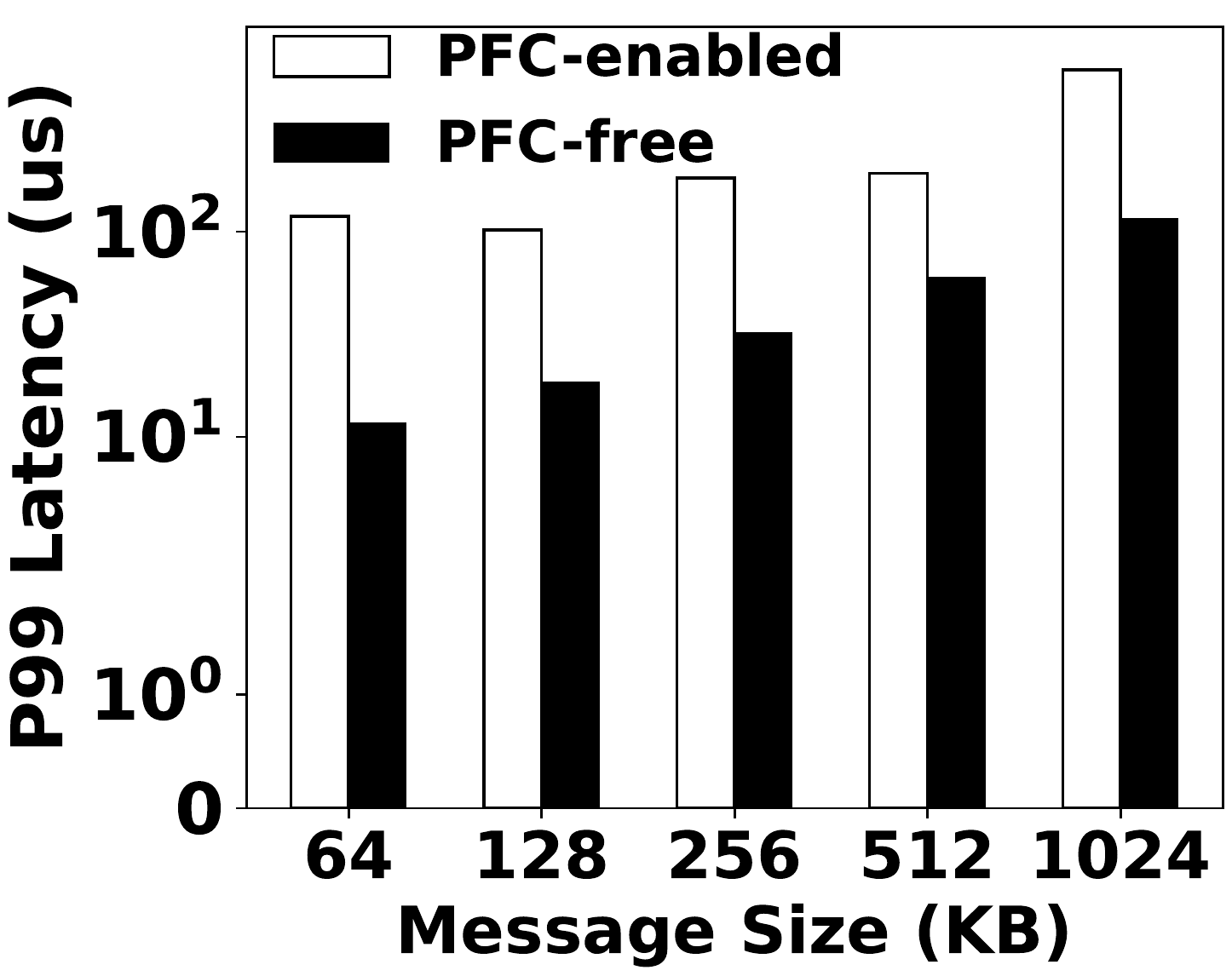}
		\caption{Network P99 latency.}
		\label{fig:moti-p99-lat}
	\end{subfigure}    
 \quad
 	\begin{subfigure}[t]{0.45\linewidth}
		\centering
		\includegraphics[width=\linewidth]{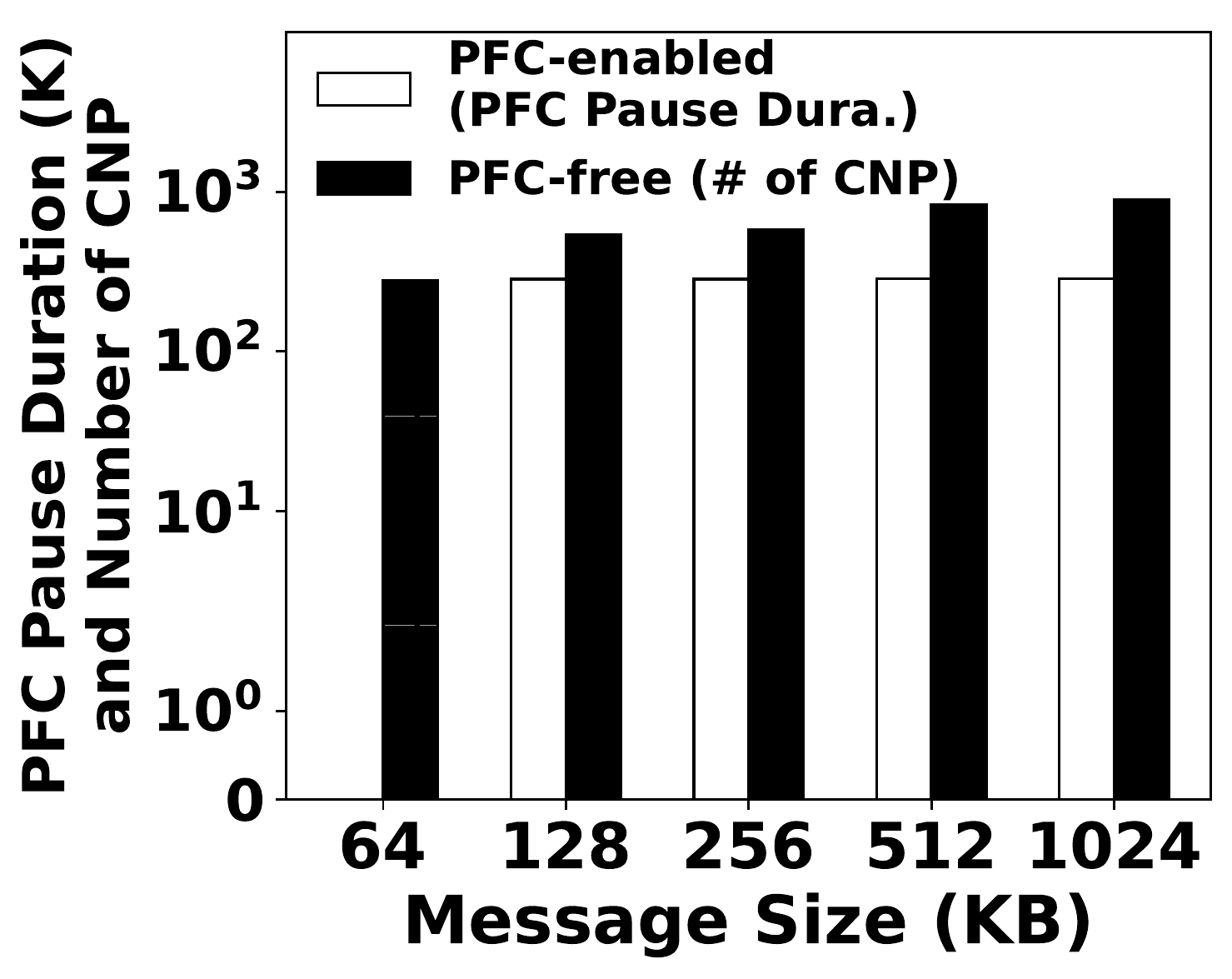}
		\caption{Average PFC and CNP.}
		\label{fig:moti-pfc-cnp}
	\end{subfigure}  
 \vspace{-0.3em}
	\caption{Testbed: network performance under memory bandwidth bottleneck.}
	\label{fig:motivation-detail-net}
\end{figure}

This throughput drop consistently happens in all sub-DCNs. 
To be concrete, we next study a pod of the
same scale in a \textit{PFC-free} sub-DCN (\ie, running only DCQCN), where each server
has a memory bandwidth capacity of 250 GB/s and a dual-port 100
Gbps ConnectX-6 DX RNIC. In another incident, Figure~\ref{fig:moti-100lossy}
shows that the server consumes a \textasciitilde230 GB/s memory bandwidth and bursts of CNPs are generated by the ECN feature supported in ConnectX-6 DX. This
feature detects the watermark of the buffer inside the RNIC and actively sends
CNPs as congestion signals to the senders to avoid packet loss caused by the buffer
overflow. The throughput drop is therefore a direct result of this large number
of CNPs. These bursts of CNPs also happen in another incident in a 25 Gbps PFC-free
sub-DCN (Figure~\ref{fig:moti-25lossy}). Although we do not have a 100
Gbps PFC-enabled sub-DCN due to its severe PFC
storm~\cite{DBLP:conf/sigcomm/ZhuEFGLLPRYZ15,DBLP:conf/nsdi/GaoLTXZPLWLYFZL21},
we believe our observation on memory bandwidth being the bottleneck also holds in such a setting. 

\para{Why receiver memory bandwidth becomes the bottleneck?}
To fully understand the reasons behind our findings, we conduct
a controlled measurement study on a small testbed with the same hardware and
software settings as in the production DCN. We find that the root cause of this
bottleneck is the high contention of memory bandwidth among RNIC and CPU
cores. Because the RNIC cannot acquire sufficient memory bandwidth to send
received messages to memory, in-flight messages queue up in the RNIC buffer and
eventually are dropped. Consequently, network congestion control is triggered, 
resulting in a substantial drop in receiver throughput
and a considerable increase in latency. Moreover, we find that instead of
alleviating this issue, the state-of-the-art direct cache access technique (\ie,
DDIO) could even worsen the situation. Specifically, we consider two
scenarios:

\begin{itemize}
\item \textbf{A 2 * 25 Gbps PFC-enabled network}: 
	two servers interconnected with a ToR switch. Each server is equipped
		with PCIe 3.0 * 8 and a dual-port 25 Gbps ConnectX-4 LX RNIC and
		enables DDIO with 4 MB cache ways. 

\item \textbf{A 2 * 100 Gbps PFC-free network}: 
	two servers interconnected with a ToR switch. Each server is equipped
		with PCIe 4.0 * 16 and a dual-port 100 Gbps ConnectX-6 DX RNIC
		and enables DDIO with 6 MB cache ways.
\end{itemize} 

\begin{figure}[t]
\centering
 	\begin{subfigure}[t]{0.45\linewidth}
		\centering
		\includegraphics[width=\linewidth]{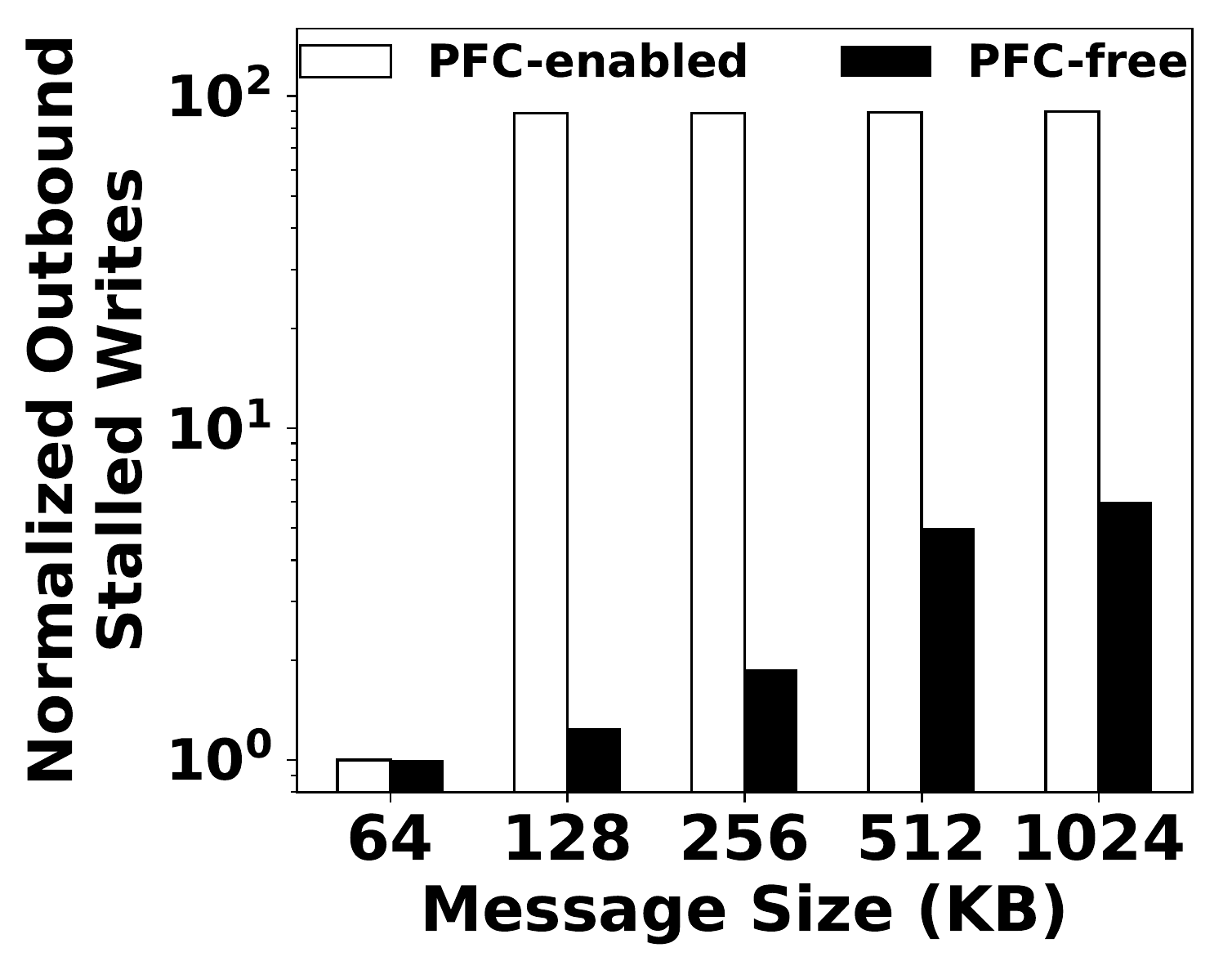}
		\caption{PCIe back pressure.}
		\label{fig:moti-pcie}
	\end{subfigure}    	
\quad
 	\begin{subfigure}[t]{0.45\linewidth}
		\centering
		\includegraphics[width=\linewidth]{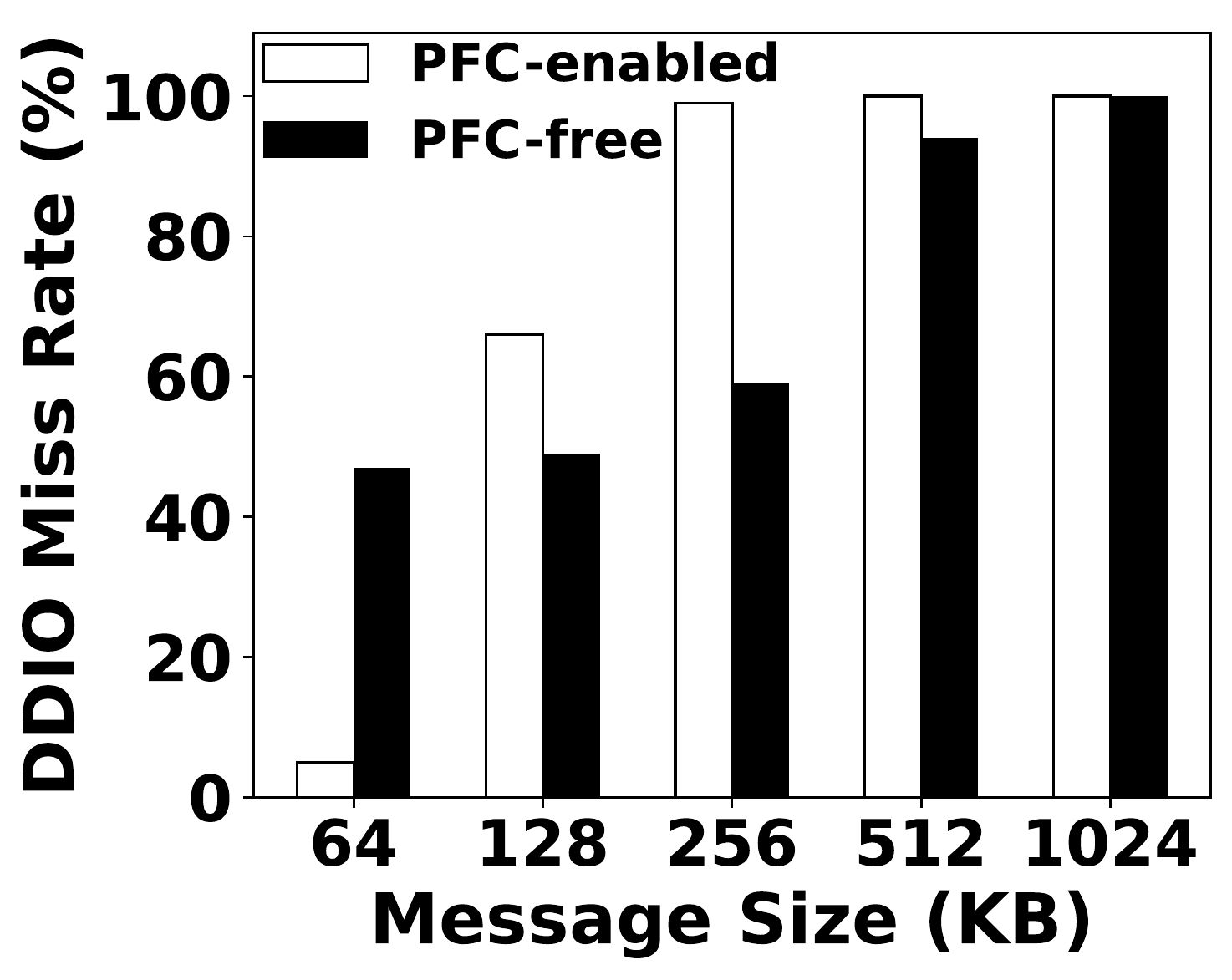}
		\caption{DDIO miss rate.}
		\label{fig:moti-ddio}
	\end{subfigure}    	
\caption{Testbed: PCIe back pressure and DDIO miss rate under memory bandwidth bottleneck.}
\label{fig:motivation-detail-nic}
\end{figure}

We let one server play the sender to send data with 32 QPs 
to another server through RDMA ib\_write\_bw. We vary the message size to change
the memory footprint of each QP.
We simulate CPU cores' memory bandwidth consumption using RDT
\textit{membw}~\cite{Intel_rdt}, which generates memory
bandwidth by reading and writing the memory of 128 MB chunks at a certain frequency.
If there is no network activity, \textit{membw} will occupy all available memory
bandwidth. 
We use NVIDIA neohost~\cite{neo-host} to measure the number of cycles
 that the PCIe unit had outbound posted write requests but could not operate. We call it the PCIe outbound stalled writes and
use it to measure the pressure PCIe brought on RNIC. 

We measure how effective DDIO is
in alleviating the receiver memory bandwidth bottleneck. We use Intel
Performance Counter Monitor (PCM) ~\cite{intel-pcm} to count the numbers of PCIe 
write hit/miss, which reflects the occurrence of DDIO write updates / allocate. 
More write updates
are desirable because it means that RNIC can frequently write data to the cache
without swapping it between cache and memory. In contrast, a higher occurrence
of write allocate indicates otherwise and hence is not
desirable~\cite{DBLP:conf/usenix/FarshinRMK20,
DBLP:conf/nsdi/TootoonchianPLW18}.


\para{Receiver throughput drops and latency increases under high memory
bandwidth contention.}
Figure~\ref{fig:motivation-detail-net} plots the receiver throughput, latency, and the corresponding PFC pause duration/CNP in our experiments. Specifically,
the receiver throughput/latency  decreases/increases as the message size
increases.  In the 1-MB message size setting, in the PFC-enabled and PFC-free
network, the throughput drops by 43.2\% and 43.1\%, respectively, to that in
the 64-KB setting.  
The increase in average and P99 latency is 11.1x/5.1x and 9.8x/8.9x, respectively.
Meanwhile, the PFC pause duration and CNP significantly increase as the message
size increases.

\para{PCIe poses severe back pressure on the RNIC.} 
Figure~\ref{fig:moti-pcie} shows that the normalized PCIe outbound stalled writes
over the 64-KB message setting increase when the message size increases.
This indicates that the high memory bandwidth contention leads to severe back
pressures from PCIe to the RNIC, which prevents the RNIC from getting sufficient
memory bandwidth and eventually causes it to drop overflowed messages. 
This back pressure is smaller in the PFC-free network than that in the
PFC-enabled network.  This is because PCIe 4.0 * 16 is more redundant for a 200
Gbps network capability and confirms that the root cause of our
observed performance degradation is insufficient memory bandwidth,
not the slow receiver problem~\cite{DBLP:conf/sigcomm/GuoWDSYPL16}.

\para{DDIO does not help under high memory bandwidth contention.}
One may expect DDIO should alleviate the high contention on memory bandwidth between
the RNIC and CPU cores because DDIO allows the RNIC to access the cache for faster
message processing. However, we find that it is not the case.
Figure~\ref{fig:moti-ddio} plots the DDIO miss rate (\ie, the frequency of DDIO
write allocate). As the contention on memory bandwidth becomes higher (\ie, the
message size increases), the DDIO miss rate also increases. When the message
size is 1 MB, the miss rate becomes 100\% in both networks. This means that
every RNIC's attempt to access the cache will trigger a write allocation, leading
to the leaky DMA problem~\cite{DBLP:conf/usenix/FarshinRMK20,
DBLP:conf/nsdi/TootoonchianPLW18}. This is because a larger message size
requires the RNIC to reserve a larger space in the cache, which, when added together,
exceeds the DDIO-allocated cache. As a result, the RNIC needs to consume an even
higher memory bandwidth to operate memory writeback when evicting cache lines
for performing DDIO write allocate. A strawman solution one may think of is to
allocate a larger LLC for DDIO. However, in our evaluation (\S\ref{sec:eval}),
we find that even if we double the LLC for DDIO, the receiver still experiences the same throughput drop and latency increase.

\subsection{RDCA: Rationale and Challenges}\label{sec:challenge}
Our proposal to cope with the memory bandwidth bottleneck is to move memory out
of the receiver host datapath, and reserve a small area in the LLC to let the
RNIC access data directly in the cache (\ie, RDCA). The rationale behind RDCA is
three-fold. First, the cache has a much higher bandwidth than the memory. 
Removing the slowest link from the datapath follows the basic principle of
optimizing system performance. 

Second, RDCA is feasible. Our study in a production DCN shows: 
after the data leaves the RNIC, no matter if it is sent to a persistent
storage media (\eg, SSD), a different processor (\eg, GPU), or consumed by
computation tasks (\eg, ranking and counting), the time it spends in memory
is usually very short (\eg, hundreds of $\mu s$ in both average and 99\%
quantile in our production DCN). As such, from Little's law~\cite{little_law}, the average
amount of data the RNIC needs to store in the memory temporally is small enough
to be held in the LLC of commodity server CPUs. For example, for an average post-RNIC timespan of 200 $\mu s$, we only need a 5 MB space on average to
support RNIC operation at 200 Gbps.


Third, RDCA is scalable. The speed of RDMA bandwidth expansion is
slower than that of the LLC capacity expansion. 
RDMA bandwidth increases 4 times in 5 years (100 Gbps/port in 2016 to 400
Gbps/port in 2021). The LLC capacity has increased over 30 times in 9 years (20
MB in Intel Xeon E5~\cite{intel_e5} in 2012 to 768 MB in AMD 3rd-gen EPYC in
2021~\cite{adm_cpu})
As such, we
foresee an efficient RDCA implementation requires no substantial
redesign in the future. 

\para{Challenges in realizing RDCA.}   
We identify three key challenges in realizing the RDCA architecture. 

\begin{itemize}
\item \textbf{The mismatch between the small size of reserved LLC and the inefficient
reservation of RDMA QPs.}
RDMA QPs usually over-reserve large buffer spaces. This leads to inefficient
statistic multiplexing of the reserved LLC and compromises the potential
benefits of RDCA. As such, we need to improve the level of statistic
multiplexing of the reserved LLC among RDMA QPs.

\item \textbf{The mismatch between reservable LLC and reserved LLC.}
Reserving too much LLC for RNIC could cause a higher
contention for the remaining LLC among other processes, impairing their
performances. As such, it is desirable to reserve a smaller area of LLC while
still supporting the RNIC processing received data at line rate. Since the line rate
is a fixed value, we need to design techniques to reduce the post-RNIC timespan
of data.

\item \textbf{The mismatch between the reserved LLC and
occasional jitters.}
The reserved LLC can hold data with an average post-RNIC timespan.  However, to
ensure that the network can consistently provide high
performance, we must take measures to cope with occasional
jitters, such as straggler data or burst arrivals of data 
with a longer post-RNIC timespan. 

\end{itemize}

Next, we present \system{}, a cache-centric receiver service
that realizes RDCA by addressing these three challenges.


\section{\system{} Overview}\label{sec:overview}
\begin{figure}[t]
\centering
\setlength{\abovecaptionskip}{0.0cm}
\setlength{\belowcaptionskip}{-0.2cm}
    \includegraphics[width=0.9\linewidth]{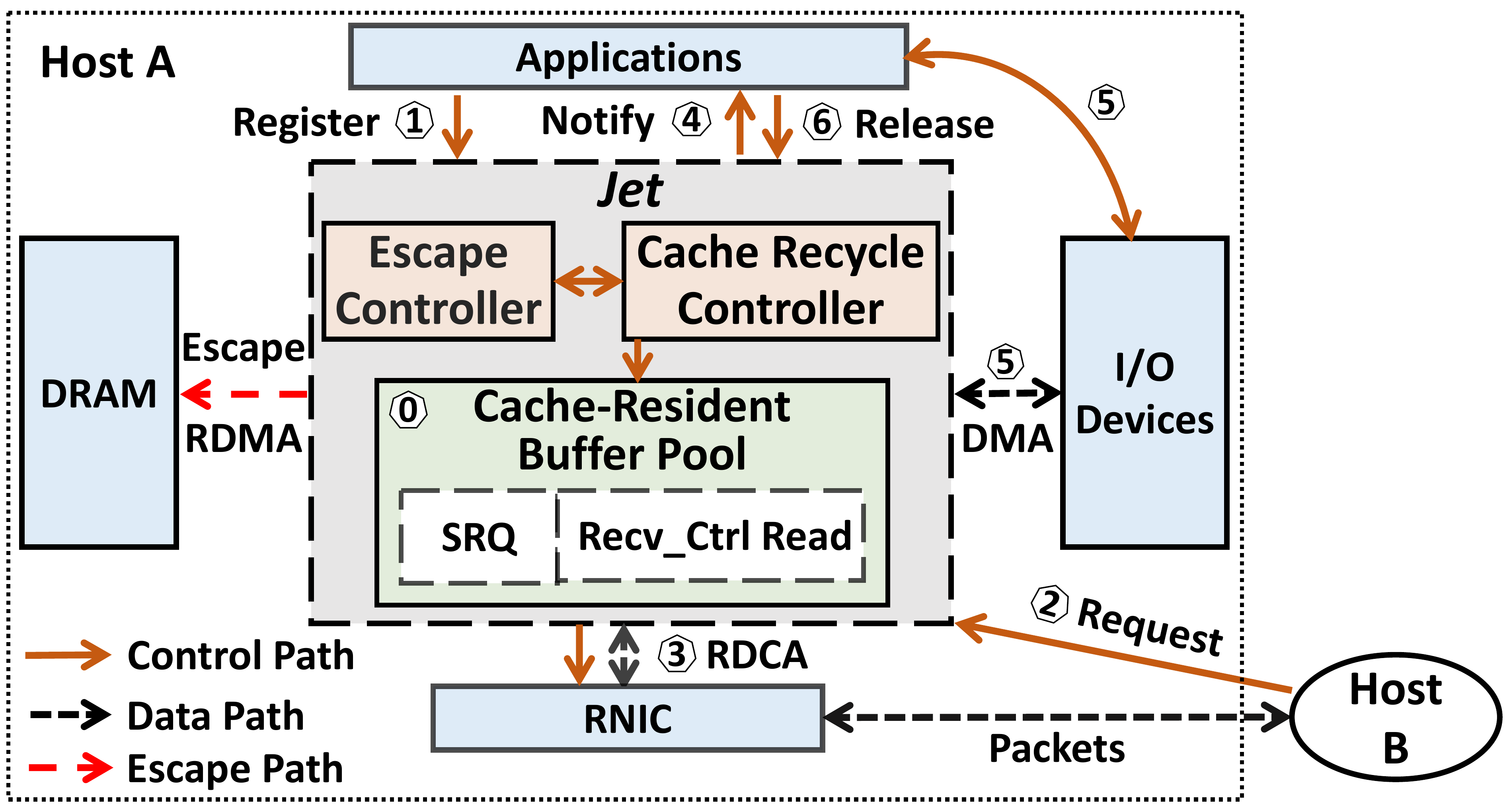}
    \caption{The basic architecture and workflow of \system{}.}
    \label{fig:workflow_v7}
\end{figure}

This section presents the architecture and the basic workflow of \system{}
(Figure~\ref{fig:workflow_v7}), a cache-centric service realizing RDCA.

\subsection{Architecture}
\system{} is a receiver-side service between applications and
the RNIC with three key components: a cache-resident buffer pool, a
cache-recycle controller and an escape controller.

\para{Cache-resident buffer pool.} 
This buffer pool is a reserved area in the LLC that
temporally stores the data leaving RNIC. Specifically, to cope with the large
space needed by different RQs, it uses an SRQ buffer to receive
small-size messages sent using RDMA SEND, and a READ buffer
equipped with a window-based rate control mechanism to receive large-size
messages sent using RDMA READ. 

\para{Cache recycle controller.}
This controller is responsible for recycling  the
cache-resident buffer to temporally store post-NIC data of different QPs.  We
develop data processing techniques such as multi-thread, pipeline, hardware
offloading and struct-based (de)serialization to reduce the post-NIC timespan
of received data, so that \system{} can multiplex a smaller reserved LLC to
support RNIC processing messages at line rate. 

\para{Escape controller.}
This controller is responsible for monitoring the usage of the cache-resident
buffer pool and taking corresponding actions to handle occasional jitters.  It
replaces the cache buffer of straggler data with a new buffer, copies the data
of slow-running applications to memory if too many replacements happened, and
lets the RNIC mark ECN in CNPs to indicate congestion if copying to
memory fails or is not sufficient in releasing the cache pressure.

\subsection{\system{} Workflow}
We illustrate the basic workflow of \system{} using an example of an application
at host A receiving data from host B and writing it to the SSD. 
Suppose host B wants to send a large chunk of data using RDMA READ. The process
is as follows.





\para{Step 0: \system{} reserves cache at initialization.}
When \system{} starts, it reserves a space in LLC as the cache-resident buffer
pool. It consists of an SRQ buffer for small-size
messages and a READ buffer for large-size
messages.

\para{Step 1: application at A registers the \system{} service.}
When the application at A wants to use the \system{} service, it first
registers with \system{}. If \system{} grants the registration, it becomes a
proxy for the application to receive data and establishes a shared cache mapping
between itself and the application. This mapping will later be used to store the
metadata of the data RNIC sends to the cache-resident
buffer pool. 

\para{Step 2: host B sends \system{} a request to transmit data using READ.}
B constructs a queue pair (QP) with \system{}.
It then sends \system{} the corresponding metadata, such as the data size and
the memory address at B. Upon receiving such a request, the cache recycle
controller adds it to an admission queue and processes receive requests in the
queue in a first-in-first-serve order. A request is granted if (1) the expected
cache footprint of the request (\ie, the expected throughput of the request
times its expected post-NIC timespan) is smaller than the remaining capacity of
the cache-resident buffer pool, and (2) the number of concurrent transmissions
in \system{} is below a predefined threshold. Once the request is admitted,
\system{} allocates an area in the READ buffer for this request.

\para{Step 3: \system{} forwards the admitted request to RNIC for data
transmission.}
After allocating the corresponding area in the cache-resident buffer pool,
\system{} forwards the request to the RNIC as well as the cached address to which
the RNIC can send the received data. The RNIC receives this request and starts
to retrieve the data from the memory of host B.

\para{Step 4: \system{} notifies the application once data arrives at
the cache.}
When the RNIC receives the data, it forwards them to the cached address in the
cache-resident buffer pool specified in Step 3.  The cache recycle controller
notifies the application process about the arrival of data with the pointer to
the corresponding address through the shared cache mapping established in
Step 0.

\para{Step 5: the application writes the received data to SSD.}
When receiving the notification from \system{}, the application sends 
SSD a pointer to the cached address of the data. SSD uses
this pointer to retrieve the data from the cache-resident buffer pool using DMA.
Step 4 and 5 continue until the whole data from B is written to the SSD
at A.

\para{Step 6: \system{} releases the allocated cache when notified by the
application.}
After finishing the data writing, SSD notifies the application, which
then notifies \system{} through the shared cache mapping. The cache recycle
controller then releases the allocated cache for further use.

If host B sends small-size messages to A using
RDMA SEND, the workflow stays the same except for Step 2 and 3: 
when constructing the QPs, \system{} posts work queue elements (WQEs) to the SRQ, binds them with a
cache area in the SRQ buffer and sends the RNIC the cached address.
Instead of sending a request first, host B directly sends the data to \system{}
and the RNIC sends the received data to the corresponding cache area.

Suppose an application does not write the received data to SSD, but instead
sends them to another processor (\eg, GPU) or performs in-memory computation
(\eg, ranking). Step 5 and 6 of the workflow also change
correspondingly.

\section{\system{} Design Details}\label{sec:design}

We now present the details on how \system{} addresses the three challenges in realizing RDCA (~\S\ref{sec:challenge}). 

\subsection{Cache-resident Buffer Pool}\label{sec:pool}

We design the cache-resident buffer pool to address the mismatch between the small size of reserved LLC
and the inefficient reservation of RDMA QPs by limiting the memory buffer used by the received data. Specifically, we adopted different RDMA operations and control methods for small and large messages respectively. For small messages, we choose the RDMA SEND/RECEIVE, together with SRQ mechanism to aggregate RQs. 
For large messages, we use RDMA READ, together with a \textit{receiver-side read control mechanism} to limit the number of concurrency and the size of in-flight messages (\eg, messages in transit).

\subsubsection{Small Messages: SEND/RECEIVE With SRQ}

\para{Using SEND/RECEIVE for low latency.}
We choose the RDMA SEND/RECEIVE operation to transmit small messages (\ie, messages smaller than 4 KB). Because small messages are latency-sensitive and require low-delay transmission. While SEDN/RECEIVE takes one RTT time less than READ when transmitting messages. However, if every QP utilizes its own RQ exclusively, it may take up a lot of memory footprint. Since in order to prepare for receiving a large number of burst messages, a sufficient number of WQEs need to be posted into the RQ~\cite{ib} in advance. These WQEs require to have MRs (Memory Regions) registered in the memory and take up memory footprint. Thus, too many WQEs would cause the size of the memory footprint to be larger than the capacity of the cache. Besides, not all the QPs are active usually, which makes the utilization of WQEs and their pointing memory footprint very low.

\para{Using SRQ to aggregate RQs.}
In order to reduce the required size of the memory footprint
, we use the SRQ~\cite{196243} to aggregate RQs. Specifically, SRQ allows multiple QPs to share one RQ, decreasing the required number of WQEs and cutting down the size of the memory footprint. When any QP receives a message, the hardware will extract a WQE from the SRQ, and store the received data according to the address provided by the WQE. Then, completion information would be returned to the corresponding upper user through the completion queue. Since SRQ can reduce the number of WQEs that need to be posted, SRQ is conducive to reducing the size of the memory footprint so that it can reside in the cache and ensure that burst messages can be handled.

\subsubsection{Large Messages: Read With Receiver-side Control}

\para{Using READ for high throughput.}
We employ the READ to transfer large messages (\ie, messages larger than 4KB). READ helps in avoiding the problem of large message failure because it allows the receiver to decide the address and length of the memory footprint allocated to receive messages. Meanwhile, after obtaining the memory address, length, and key of the remote data, the receiver can directly acquire access to the data in the sender. Therefore, the READ does not require the involvement of the sender software stack and is a unilateral operation, providing high throughput.

\para{Message fragmentation.}
We also fragment large data messages, because the amount of data sent in a single time is unfavorable for flow control. In addition, if the message is lost, the retransmission requires non-negligible overhead. 
Besides, large messages will be divided into slices in the network transmission. Too many slices are prone to cause the packets out-of-order problems, and arrived packets need to wait for the latest packet and take up the memory footprint. Thus, we slice the messages into no larger than 256 KB fragments before transmitting them.



\para{Receiver-side control mechanism.}
To restrict the memory footprint used by READ, we design the \textit{receiver-side read control} mechanism~\cite{montazeri2018homa,10.1145/3387514.3405897}, which consists of two parts: concurrency number control, and in-flight buffer size control.

\begin{figure}[t]
\centering
    \includegraphics[width=0.8\linewidth]{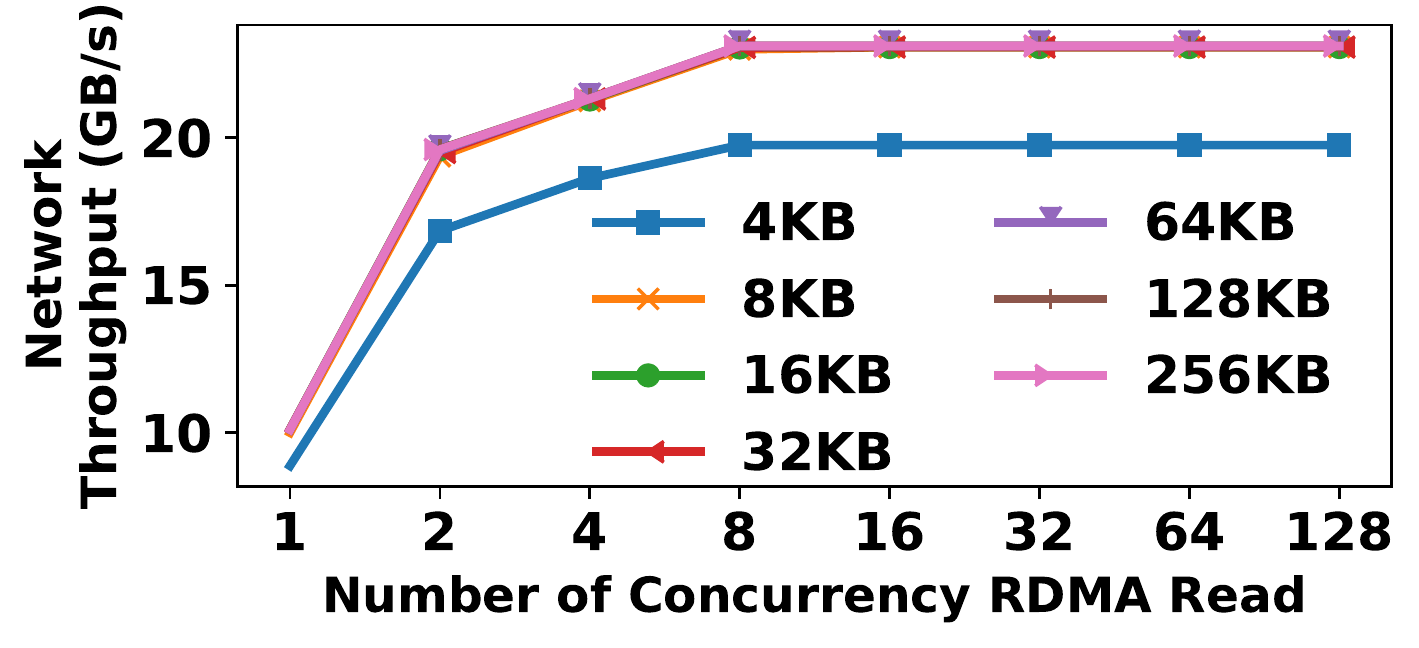}
    \caption{The network throughput of different READ concurrency numbers and message sizes.}
    \label{fig:read_concurrency}
\end{figure}

\para{1. The concurrency number control.}
A fixed-size sliding window is maintained to limit the number of admitted READ requests. Only requests that are in the window are allowed to be sent. If there is not enough window capacity, the requests will be waiting in the request queue, which is a first-in-first-out queue. Until the previous requests are processed, the window moves, and then the subsequent requests can be processed. With a fixed-size sliding window, we can control the number of concurrent READ requests to be less than a certain value (\eg, 32). What's more, this concurrency number control can also help to solve the in-cast problem, which is caused by burst messages arriving at the receiver RNIC simultaneously and would affect the latency and throughput of message transmission~\cite{10.1145/3387514.3405897}. 

Besides, we do experiments to test how many concurrencies and how large the message size can fully utilize the network bandwidth using a single thread. 
As the results shown in Figure~\ref{fig:read_concurrency}, when the concurrency number equals to four, the 2*100 Gbps network bandwidth can be fully utilized. Usually, the number of active network connections is 32. Thus, we set the concurrency number 32. 

\para{2. The in-flight buffer size control.}
We use another window to limit the total amount of memory footprint taken up by in-flight READ data. Specifically, a fixed-size window (e.g., 8MB) is maintained to bound the dynamic number of inbound bytes. If the available window capacity is less than the size of messages requested by READ, this READ request would be queued and deferred until sufficient window capacity is allocated to the request. When a READ request is processed, its corresponding occupied window capacity can be released for subsequent read requests.

\subsubsection{Deciding the size of the cache-resident buffer pool}
To ensure that the capacity of the cache-resident buffer pool can meet the requirements of receiving data and is not too large, we take both the message size and concurrency number into consideration carefully to calculate its size. Specifically, for 32 QPs we prepare 1K WQEs in advance in the SRQ to deal with received small messages, and the size of each WQE is 4 KB. Thus, the size of the SRQ buffer is initially set as 4 MB. Considering that the size of the largest message is 256 KB and the READ concurrency number is 32, the READ buffer allocated for big messages is initialized as 8 MB.
Note that the SRQ buffer and the READ buffer share the whole 12 MB space. Thus, the size allocation between the two buffers can be dynamically adjusted at runtime, only guaranteeing that the size of the SRQ buffer is larger than a threshold.

\subsection{Swift Cache Recycle}\label{sec:recycle}

In order to eliminate the mismatch between reservable LLC and reserved LLC, we design a \textit{swift cache recycle mechanism} in the cache recycle controller to reduce the required cache size and shorten the time that applications occupy the cache buffer. Specifically, first, we take advantage of the slab~\cite{252795} algorithm to manage the cache buffer in the granularity of 4 KB. Because the slab algorithm can divide the cache buffer into object pools, solving the problem of internal fragmentation. 
Second, \system{} and applications transfer metadata and release requests of cache buffers through shared caches to realize cache buffer recycle. Third, in order to speed up the process of the cache cycle, we have made the following optimizations: (1) use multi-thread to perform data processing operations at the same time and achieve parallelism, (2) leverage pipeline technology to realize cache recycle in finer granularity, and (3) simplify the data process to further improve efficiency and shorten the time.

\subsubsection{Cache Buffer Recycle}

To be specific, in the initialization phase, \textit{shmget} is used to allocate one cache space shared by \system{} and the application in the cache. Also, \textit{shmat} is used to mount and map this shared space to both \system{} and the application address spaces. In this way, after getting the received data from RNIC to the cache buffer in the cache-resident buffer pool, \system{} delivers the address pointer pointing to the cache buffer to the application through shared cache. If the data needs to be further transmitted to the I/O devices, the I/O devices can obtain the data through DMA operation. When the application processes finish receiving and processing the data, they can use the shared cache again to inform \system{} to release the cache buffer. The occupied cache buffer is released back to the cache-resident buffer pool, which can be used for the subsequent requests, realizing the cache buffer recycle. 

\subsubsection{Accelerating the Recycle}

We accelerate the cache buffer reuse rate to guarantee the availability of \system{} and shorten the timespan data spent in the cache.
The relationship between \system{} and applications can be abstracted as the producer-consumer. \system{} is responsible for getting the received data into the cache buffer, which plays the role of the producer. While the applications are responsible for using the obtained data from the cache buffer, thus applications can be regarded as consumers.
Thus, in order to recycle the cache more quickly, we speed up the rate at which producers produce data through three methods: using multi-thread to achieve data processing parallelism, segmenting data to form a pipeline, and simplifying data processing.

\para{Realizing processing parallelism through multi-thread.}
Due to the differences between the network software stack and application software stack, data packets received by the RNIC need to be processed by \system{} first. For example, there are cyclic redundancy check (CRC) checksum, data serialization/deserialization, and data formation. These non-negligible data processing overheads will slow down the speed at which \system{} provides the data to the applications. Therefore, in order to narrow the gap between the speed of data produced by \system{} and the speed of data consumed by the application, we introduce multi-thread. These threads can perform data processing operations concurrently to achieve parallelism and shorten the processing time. 

\para{Forming a data processing pipeline.}
We use I/O pipeline processing to match the rate between producer and consumer in a finer-grained. Specifically, the timespan in data processing can be roughly classified into three stages: get, process, and release. First, \system{} gets the data into the cache-resident buffer pool from the RNIC. Then, \system{} processes the data as needed to meet the requirements of the applications. At last, \system{} forwards the data to applications and applications notify \system{} to release the cache buffer. In order to form a data processing pipeline, we segment the data messages into slices no larger than 4 KB~\cite{10.1145/3387514.3405897}. These slices enter the three stages sequentially and continuously, and once the timespan of a slice finishes, the cache buffer it occupied can be released immediately. Thus, there is no need to wait until the application receives the whole message and notifies to release the cache buffer. By forming this small-slice data processing pipeline, we can recycle the cache-resident buffer pool more efficiently. 

\para{Simplifying the data processing procedure.}
What's more, we simplify the processing software and offer a simpler way to (de)serialize data as well. First, some data processing tasks can be offloaded to dedicated hardware. Take the storage system as an example. The CRC calculation is offloaded to an RNIC with CRC check calculation capabilities (\eg, Mellanox CX-5\cite{mellanoxcx-5}), so that \system{} is not required to calculate the CRC and thus save data processing time~\cite{DBLP:conf/nsdi/GaoLTXZPLWLYFZL21}. 
Second, we provide a compiler that can (de)serialize data for remote procedure call using huibuffer\cite{DBLP:conf/usenix/LiYDMLH20}, a lightweight data serialization protocol. Specifically, serialization and deserialization are data conversion technologies. Serialization is to reorganize the data according to the specified format while keeping the original data semantics unchanged. Deserialization is to restore the serialized data to semantic data and regain the original data. Unlike protobuf~\cite{protobuf} which needs memory copy and uses objects in (de)serialization, huibuffer can achieve in-place (de)serialization and use struct. Huibuffer has the advantages of flexibility and high performance.
\subsection{Cache Pressure-Aware Escape Mechanism}\label{sec:escape}

\begin{algorithm}[t]
\caption{The escape algorithm.}\label{alg:escape}
\small

\Fn{escape()}{
\textcolor{blue}{// The cache pool is insufficient. }\\
\eIf{\textit{avl\_cache\_pool} < CACHE\_SAFE}{
\textcolor{blue}{// The memory size for escape is small.}\\
\eIf{replace\_mem\_size < MEM\_ESC}{
\textcolor{blue}{// Replace straggle buffers.}\\
buffer\_replace(); }
{\textcolor{blue}{// The escape memory size is large, the data of slow-release applications needs to be copied to memory.}\\
\ForEach{app\_id}{
\textcolor{blue}{// Most cache buffer of \textit{app\_id} is straggler.}\\
\If{\large{$\frac{\textit{straggler\_buf\_num\_id}}{\textit{held\_buf\_num\_id}}$} > \small{CREDIT}}{
\textcolor{blue}{// Copy data to memory by multi-thread.}\\
data\_copy(\textit{app\_id}); 
}}}
\textcolor{blue}{// Data copy does not work because of exceptions.}\\
\If{\textit{avl\_cache\_pool} < CACHE\_DANGER}{
mark\_ecn();
}
}
{
\textcolor{blue}{// The cache pool is enough, do not need to escape.}\\
Do nothing;}
}


\end{algorithm}

An escape mechanism, which perceives the cache pressure and adjusts actively, is designed to deal with the mismatch between the reserved LLC and occasional jitters (\eg, SSD slow write, and application exceptions). Specifically, when there is not enough capacity for the slab to allocate buffers in the cache-resident buffer pool, it means the pressure of the cache is too high. In this case, the escape controller will trigger this mechanism as a safety net. Our evaluation (\S\ref{sec:eval}) shows that the probability of it being triggered is very low.

\para{Escape algorithm.}
The escape algorithm is summarized in Algorithm \ref{alg:escape}. 
There are five parameters in this algorithm: \textit{CACHE\_SAFE} and \textit{CACHE\_DANGER} measures the usage of the cache-resident buffer pool usage, \textit{MEM\_ESC} measure the size of memory used for escape, and \textit{CREDIT} measures whether an application is slowly-releasing. Note that the \textit{CACHE\_DANGER} is lower than the \textit{CACHE\_SAFE}. 
Besides, there are two main variables. \textit{held\_buf\_num\_id} represents the total number of buffers occupied by application numbered \textit{id}. 
If the time an application occupies the cache buffer is greater than a time threshold, the cache buffer will be considered a straggler buffer. The number of straggler buffers occupied by application numbered \textit{id} is recorded as \textit{straggler\_buf\_num\_id}. For an application numbered \textit{id}, if the value of \textit{straggler\_buf\_num\_id} divided by \textit{held\_buf\_num\_id} is greater than \textit{CREDIT}, the application is considered to be a slowly-releasing one.
To add, the cache buffer allocated for each application is organized by linked lists. If a new cache buffer is allocated to an application, then a node would be linked to the tail of its linked list. Thus, the timestamp of the cache buffers in a linked list is monotonically increasing. Thus, we can know whether the time of a cache buffer exceeds the time threshold by checking the timestamp of the head node in the linked list. This method provides a time complexity of O(1).





In this escape algorithm, we design three different escape actions to cope with three different degrees of abnormalities. To be specific, when the available size of the cache-resident buffer pool is smaller than \textit{CACHE\_SAFE} and the used memory size for escape is smaller than \textit{MEM
\_ESC}, we replace the straggler buffers. Else, for each application whose straggler cache buffers take up more than \textit{CREDIT} percent of total occupied cache buffers, we copy the data belonging to the application from the cache into the memory. The impact caused by copying data to memory on the memory bandwidth is negligible. Since using $\alpha$ to represent the proportion of influenced flows when escape happens, the added memory bandwidth consumption equals at most to $\alpha \times network\_bandwidth $. Considering the extreme situation where the network throughput is 100 Gbps and $\alpha$ is 4\%, the memory bandwidth only increases at most by 1 GB/s.

However, copying data from the cache to memory may do not work due to some exceptions, such as machine
check exception~\cite{conf/sigcomm/MiaoZMQZLCGZZLS22}) which is the hardware error in the processor, cache, memory, or system bus. In this scene, the available size of the cache-resident buffer pool becomes lower than \textit{CACHE\_DANGER}. To solve this problem, we mark ECN in CNPs to indicate congestion in the receiver datapath and slow down the data-sending rate.

\para{Replacing straggler buffers.}
In this escape action, we simply do not use the straggler buffer for cache recycling. Instead, an extra buffer of the same size is allocated in the memory and pre-touched, joining the cache-resident buffer pool for recycling. In this way, the total size of the usable and recyclable cache-resident buffer pool remains unchanged. A background task is responsible for replacing the straggler buffer which reads the memory footprint sequentially and periodically at a low frequency.

\para{Copying the data to the memory.}
In this escape action, we copy all the data in cache buffers occupied by slowly-releasing applications into the memory. By doing this, the straggler cache buffer can be released quickly. Copying the data to the memory will not consume extra memory bandwidth (since using RDMA also needs memory bandwidth), and can guarantee that the packets received from the network would not be dropped due to memory bandwidth bottleneck. Because copying the data from cache to memory is initiated by the CPU instead of RNIC, and in the allocation of memory bandwidth resources, the CPU has higher priority than the RNIC. In this way, we can fully utilize the network bandwidth which is more precious than the memory bandwidth.

\para{Marking ECN.} 
The enhancement of hardware capabilities enables the RNIC to handle more complex tasks and functions, including marking ECN in CNPs. Although marking ECN is a new function of the latest RNICs (\eg, ConnectX-6 DX \cite{connectx-6dx}), previous models of RNICs can also implement such a function by adding firmware. Thus, we mark ECNs to indicate the available size of the cache-resident buffer pool is below \textit{CACHE\_DANGER}. By doing this, the sender would slow down the speed of sending data when receiving CNPs, alleviating the pressure on the cache on the receiver. In our experiments, this mechanism is barely triggered.

\section{Implementation}\label{sec:impl}
We implement a prototype of \system{} in ~7K C code and deploy it in commodity
servers. We encounter three practical issues. 

\para{Cache isolation: a problem with CAT.}
We use the Cache Allocation Technology (CAT)~\cite{cat} to reserve a dedicated
LLC for \system{}.  However, we find that data in the reserved cache can be
swapped out even when the RNIC is not receiving data at line rate. We identify
two reasons. First, both the code of \system{} and the host network stack
compete for the reserved cache. As such, the data could be forced out by the
code. We address this by slightly increasing the reserved cache size and 
periodically touch random data in the buffer pool. Second, scheduling mechanisms
in operating systems (\eg, cache prefetching) could cause other data to occupy
\system{} reserved cache. As such, we turn off these functions during deployment.
In our evaluation, we find that doing so does not affect the overall application
performances.

\para{Marking ECN without corresponding RNIC interfaces.}
The last resort of the escape controller is to let RNIC mark ECNs in
CNPs. However, the ConnectX
RNICs on our servers do not provide interfaces to send CNPs. We
workaround this limitation by using a provided interface to adjust the ECN 
marking threshold of the RNIC: lower the threshold when the cache buffer is
being exhausted so that the RNIC can start to send CNPs, and 
resume it when the available cache buffer is above the safety line. Although the
adjusting ECN threshold takes time and is a seldom taken action in daily
DCN operation, we do not expect to activate this action frequently in \system{},
but treat it as a safe net. In fact, in our evaluation, we observe very rare 
activation of this action.

\para{Guaranteeing application QoS.}
We implement admission queues with different priorities.
Applications with higher QoS level is admitted access to \system{} cache pool
with a higher priority. If there is no available cache buffer for an application
with low QoS level, \system{} will let them use memory buffer instead.

\section{Performance Evaluation}
\label{sec:eval}
We evaluate \system{} in different networks that differ in scale, RNICs, CPUs and workloads.
We compare \system{} and the default data packet processing design~(DDIO) in the experiments. We aim to evaluate the effectiveness of all designs in \system{}~(\S~\ref{subsec:eval-micro}), the application-level effectiveness~(\S~\ref{subsec:eval-small-tesetbed}), and the generality of \system{}.~(\S~\ref{subsec:eval-dnn})

\subsection{Experimental Setup}
\label{subsec:eval-micro}
\para{Testbed setup.} The testbed contains two hosts: one client and one server. Both hosts are deployed with \system{}. We use two testbeds in~\S~\ref{sec:measurement}: 
the 2 * 25~Gbps PFC-enabled network and the 2 * 100~Gbps PFC-free network. We compare \system{} with DDIO. For fairness, we allocate 12 MB LLC for DDIO and \system{}~(4 MB for small messages and 8 MB for large messages~(\S\ref{sec:pool})) in both networks. We run a synthetic storage benchmark provided by a major cloud  provider, which simulates the process of its production DCN except that data will not be written into the SSDs. We configure the tool to produce messages with sizes between 4~KB and 256~KB.

\para{Production DCN.} We conduct experiments on the disaggregated storage architecture, which includes a storage cluster and a computing cluster. The storage system consists of 16 hosts. For high availability, each host is configured with a dual-port Mellanox RNIC connected to 2 ToR switches. These ToR switches are connected through spine switches. The topology is a typical Clos network, which is widely adopted in real data centers. Note that we run RoCE RDMA without TCP. In this cluster, we use the same configured hosts as the second testbed in \S\ref{sec:measurement}. Each host is equipped with 12 SSD and 2 Optane SSD. In this cluster, we also configure  \system{} with 4 MB LLC for small messages and 8 MB LLC for large messages~(\S\ref{sec:pool}). For fairness,  DDIO also uses 12 MB LLC. The computing cluster has another 16 hosts. These hosts send three sizes of messages: 4~KB, 16~KB, and 256~KB, where the messages are produced with the modified FIO~\cite{fio}, which is a program that can set flexible workloads for I/O performance tests. We randomly select a host to measure its performance, given that the performance is similar across hosts.

\subsection{Testbed Micro-benchmark Experiments}
\begin{figure*}[t]
	\centering
 	\begin{subfigure}[t]{0.18\linewidth}
		\centering
		\includegraphics[width=\linewidth]{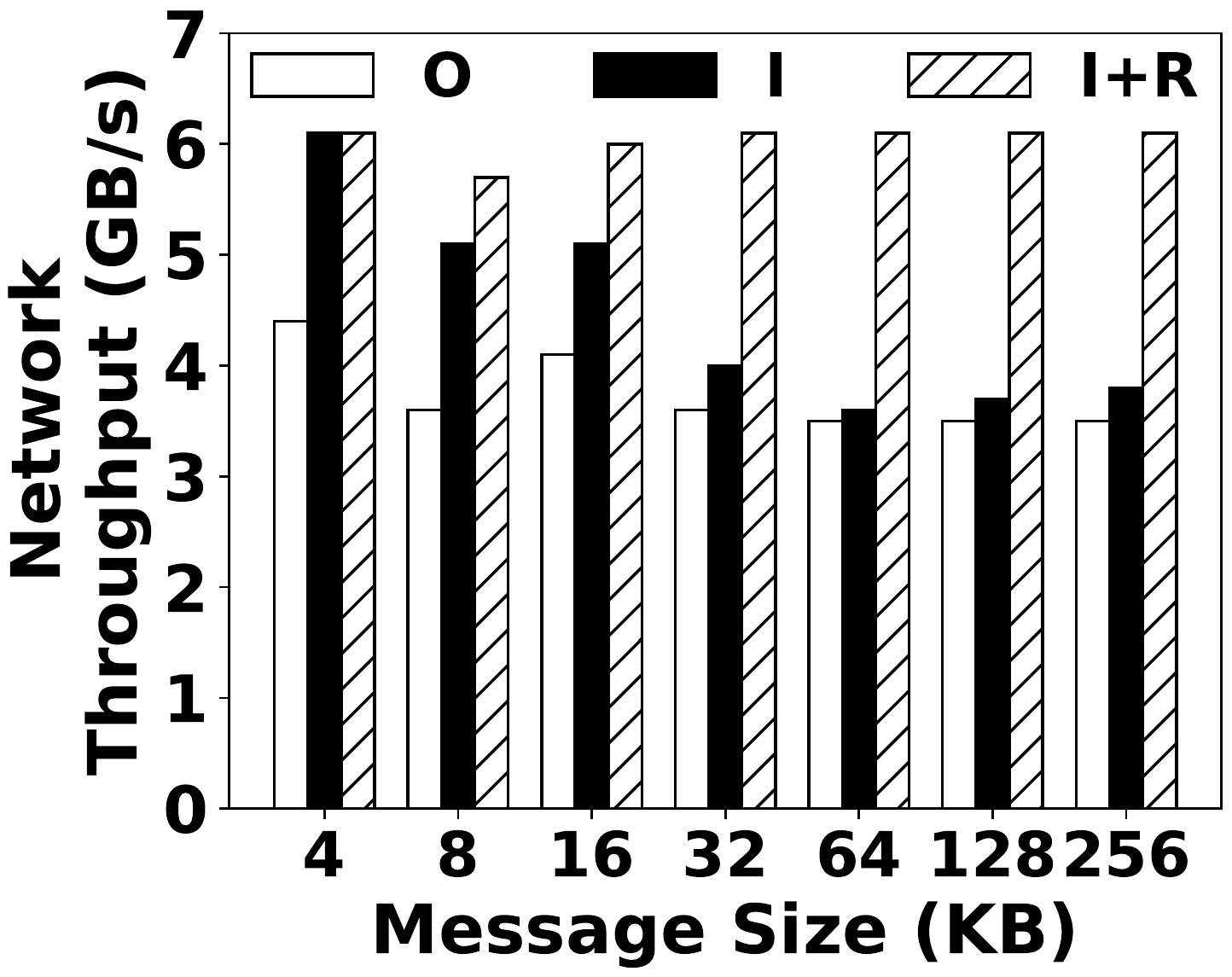}
		\caption{Throughput.}
		\label{fig:eval-micro-25-bandwidth}
	\end{subfigure}
\quad
 	\begin{subfigure}[t]{0.18\linewidth}
		\centering
		\includegraphics[width=\linewidth]{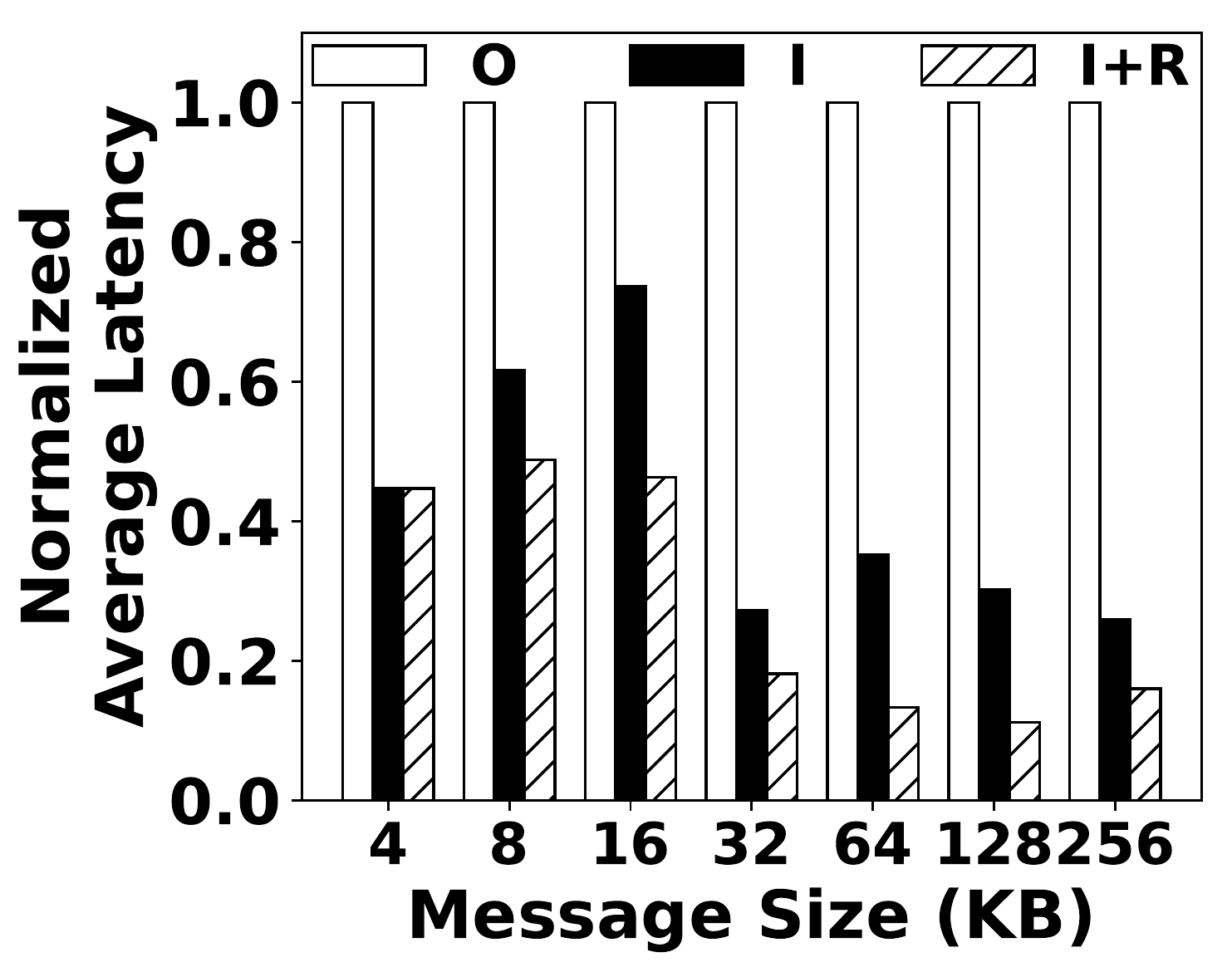}
		\caption{Average latency.}
		\label{fig:eval-micro-25-latency}
	\end{subfigure}
\quad
 	\begin{subfigure}[t]{0.18\linewidth}
		\centering
		\includegraphics[width=\linewidth]{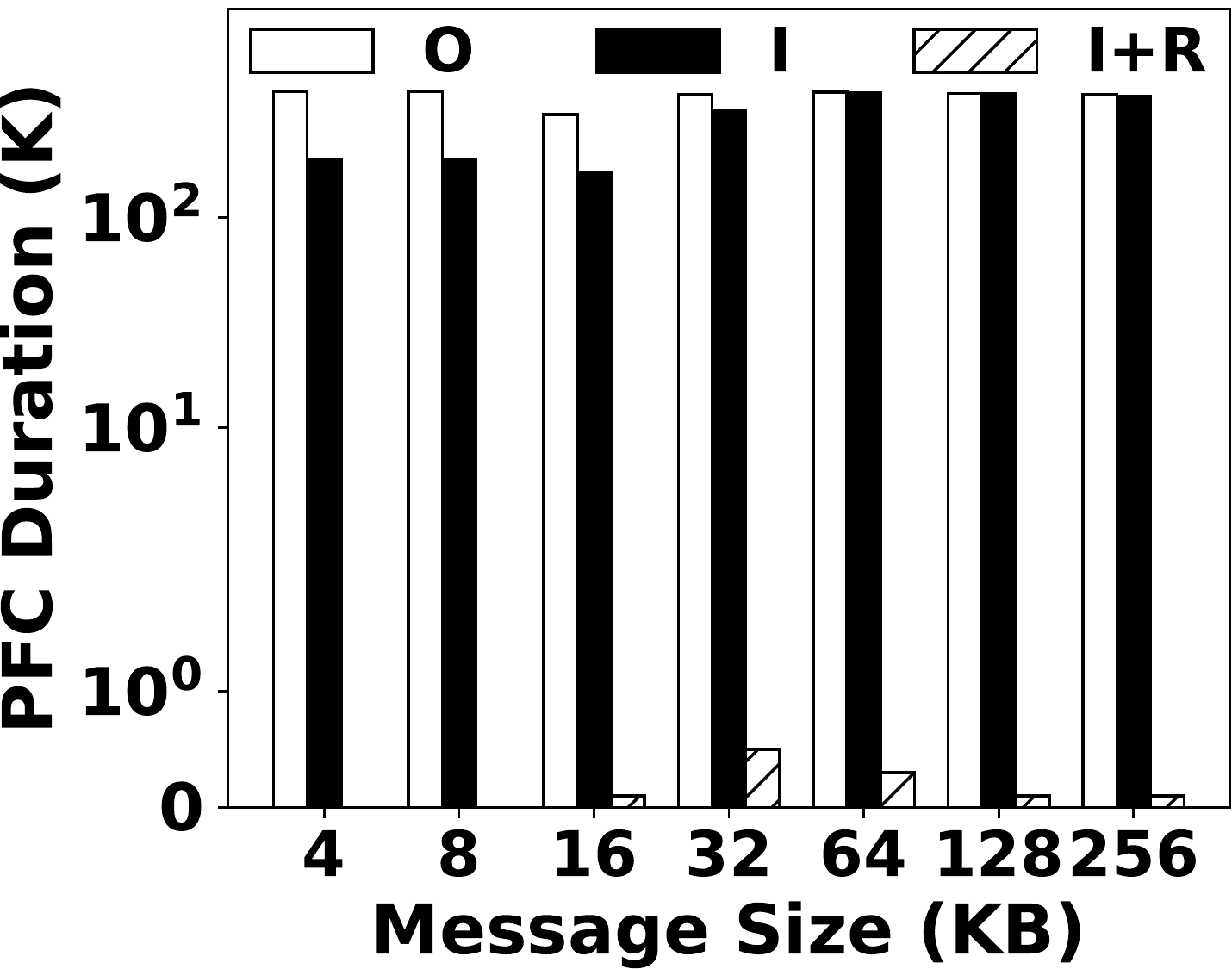}
		\caption{PFC.}
		\label{fig:eval-micro-25-pfc}
	\end{subfigure}
\quad
 	\begin{subfigure}[t]{0.18\linewidth}
		\centering
		\includegraphics[width=\linewidth]{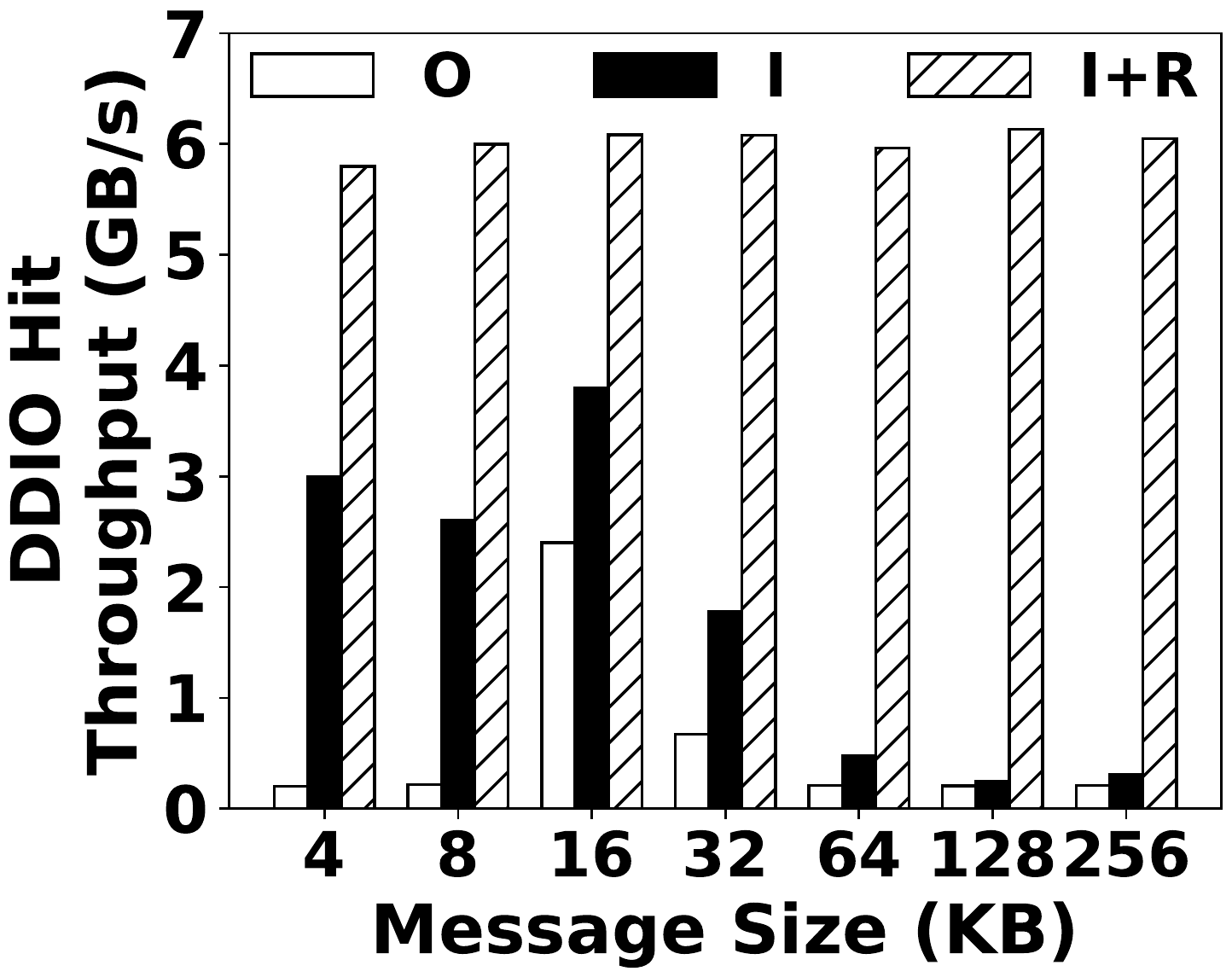}
		\caption{DDIO hit throughput.}
		\label{fig:eval-micro-25-hit-throughput}
	\end{subfigure}
\quad
 	\begin{subfigure}[t]{0.18\linewidth}
		\centering
		\includegraphics[width=\linewidth]{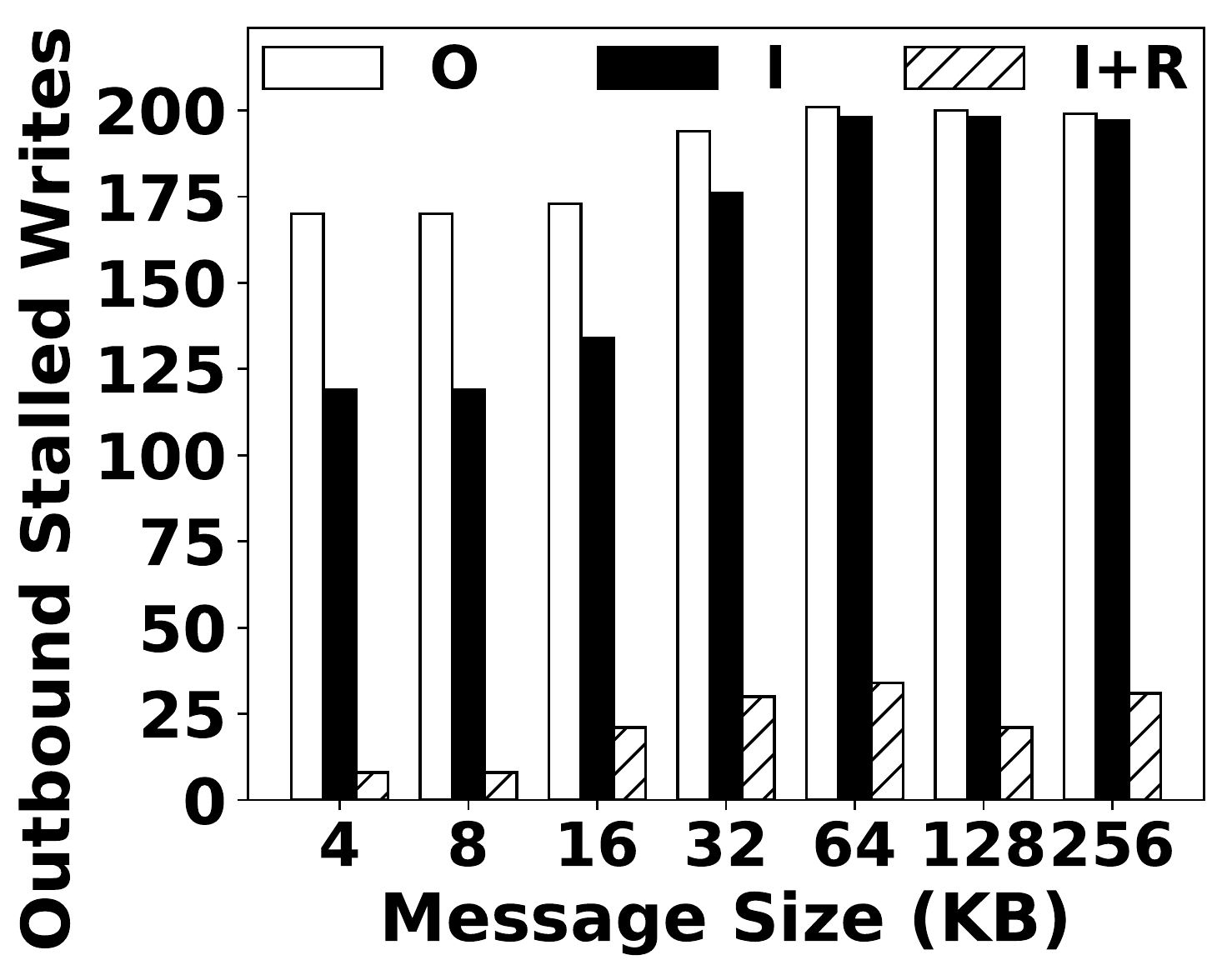}
		\caption{PCIe back pressure.}
		\label{fig:eval-micro-25-pcie}
	\end{subfigure}	
	\caption{Micro-benchmark experiment results on the dual-port 25~Gbps  PFC-enabled RDMA network. In this figure and the following figures, `O' indicates the DDIO's performances, and `I' and `R' means the performance with the cache isolation  and the cache-resident buffer pool, respectively.}
	\label{fig:group-25-micro}
 \vspace{-1em}
\end{figure*}

\begin{figure*}[t]
	\centering
  	\begin{subfigure}[t]{0.18\linewidth}
		\centering
		\includegraphics[width=\linewidth]{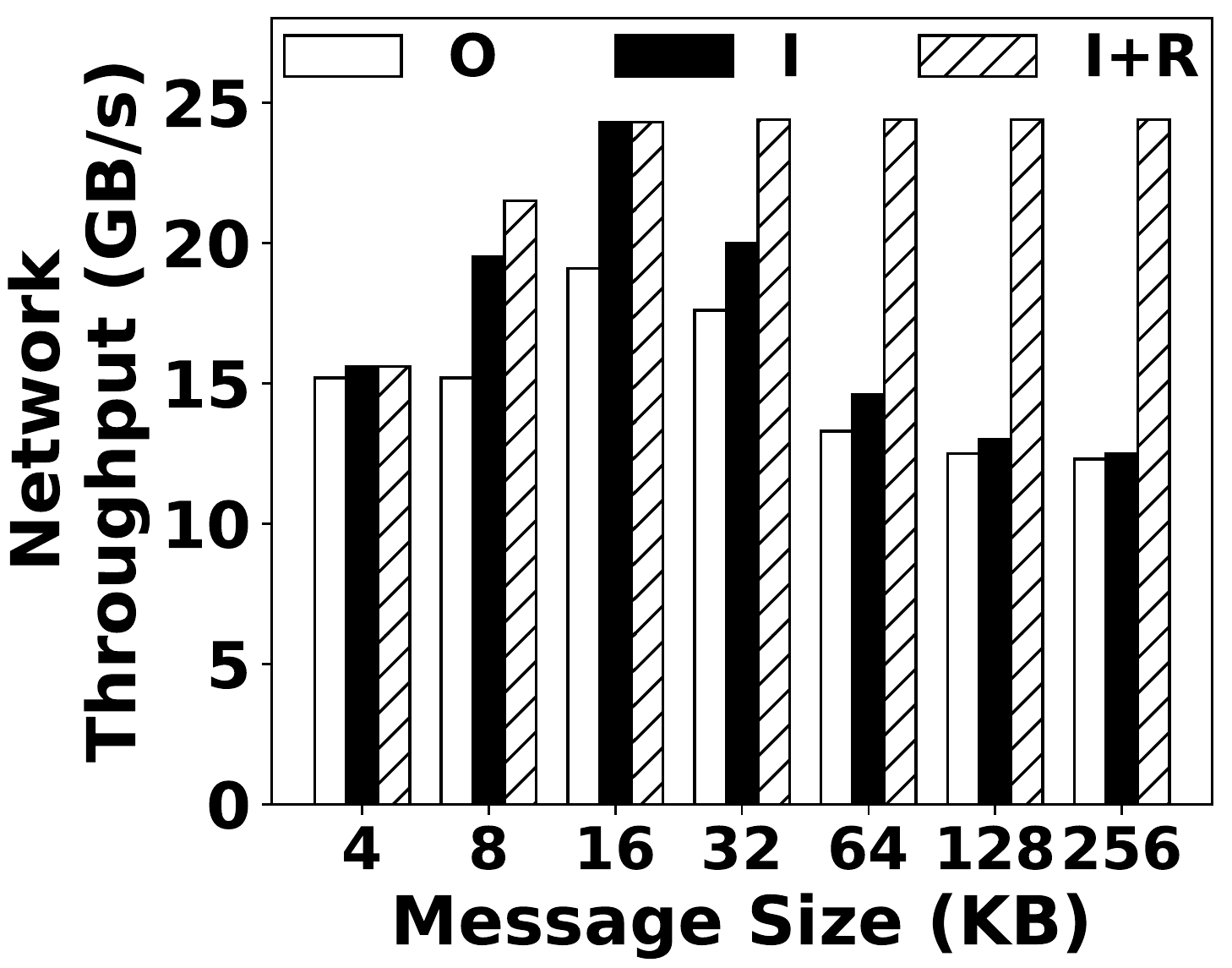}
		\caption{Throughput.}
		\label{fig:eval-micro-100-bandwidth}
	\end{subfigure}
\quad
  	\begin{subfigure}[t]{0.18\linewidth}
		\centering
		\includegraphics[width=\linewidth]{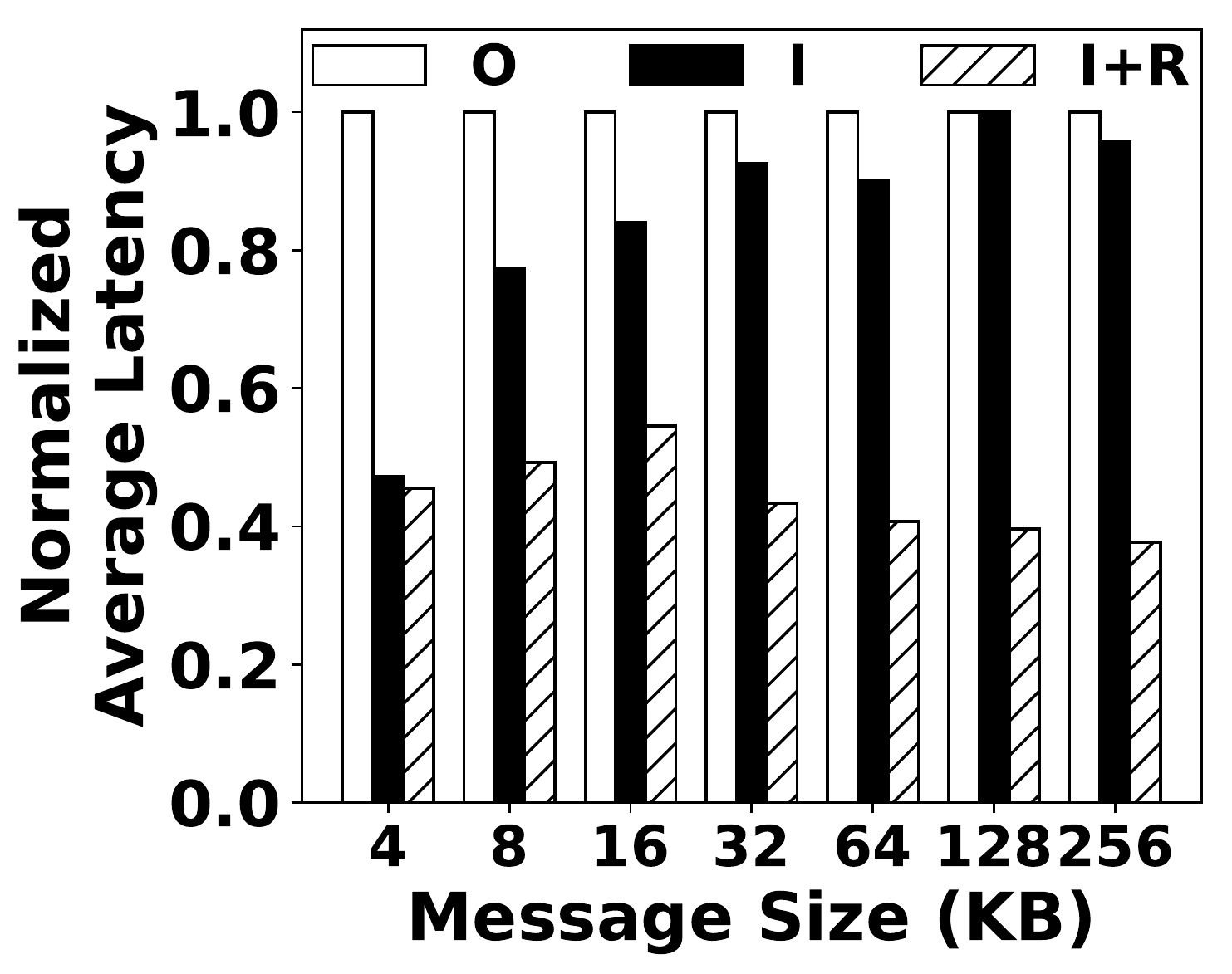}
		\caption{Average latency.}
		\label{fig:eval-micro-100-latency}
	\end{subfigure}
\quad
  	\begin{subfigure}[t]{0.18\linewidth}
		\centering
		\includegraphics[width=\linewidth]{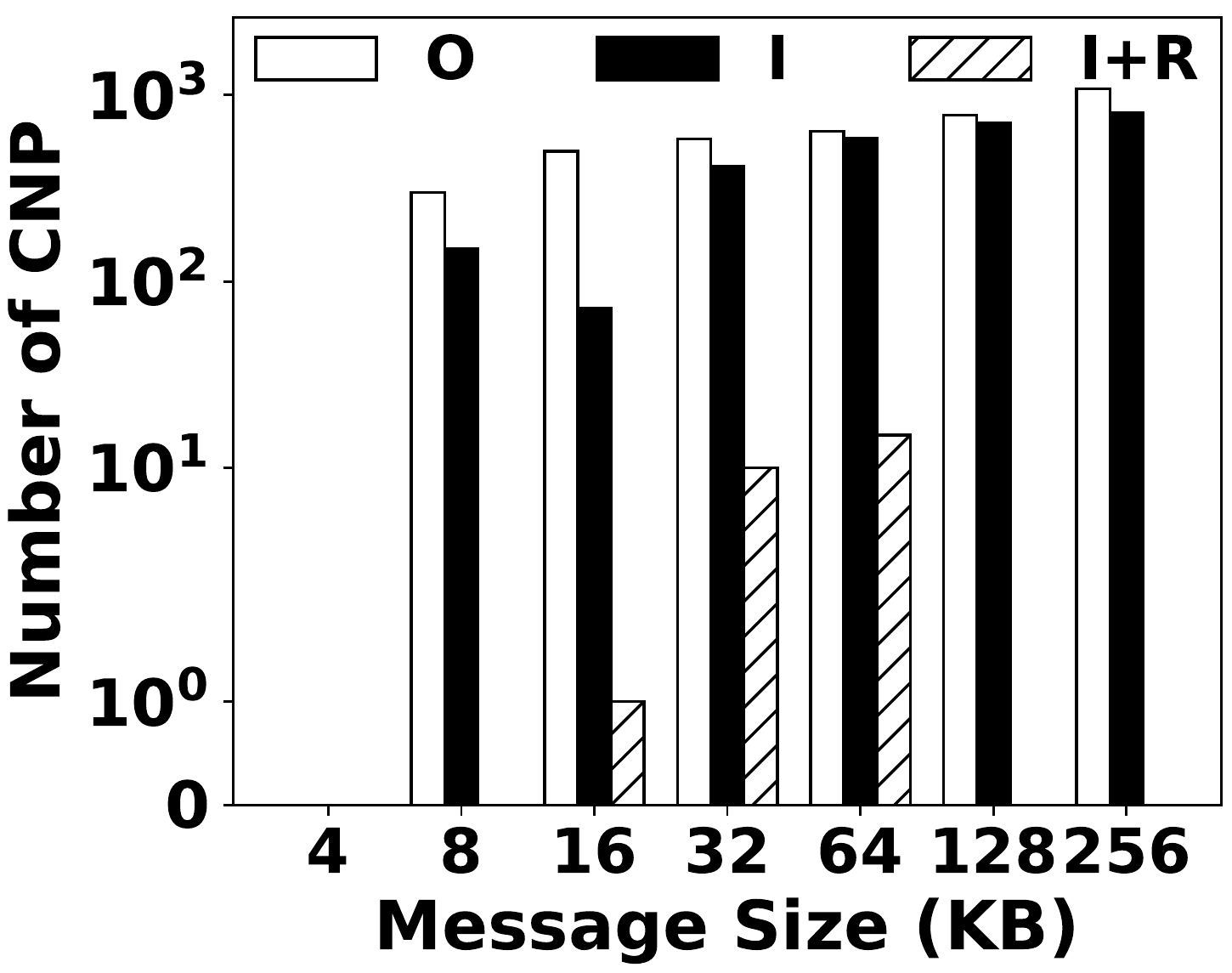}
		\caption{CNP.}
		\label{fig:eval-micro-100-cnp}
	\end{subfigure}
\quad
  	\begin{subfigure}[t]{0.18\linewidth}
		\centering
		\includegraphics[width=\linewidth]{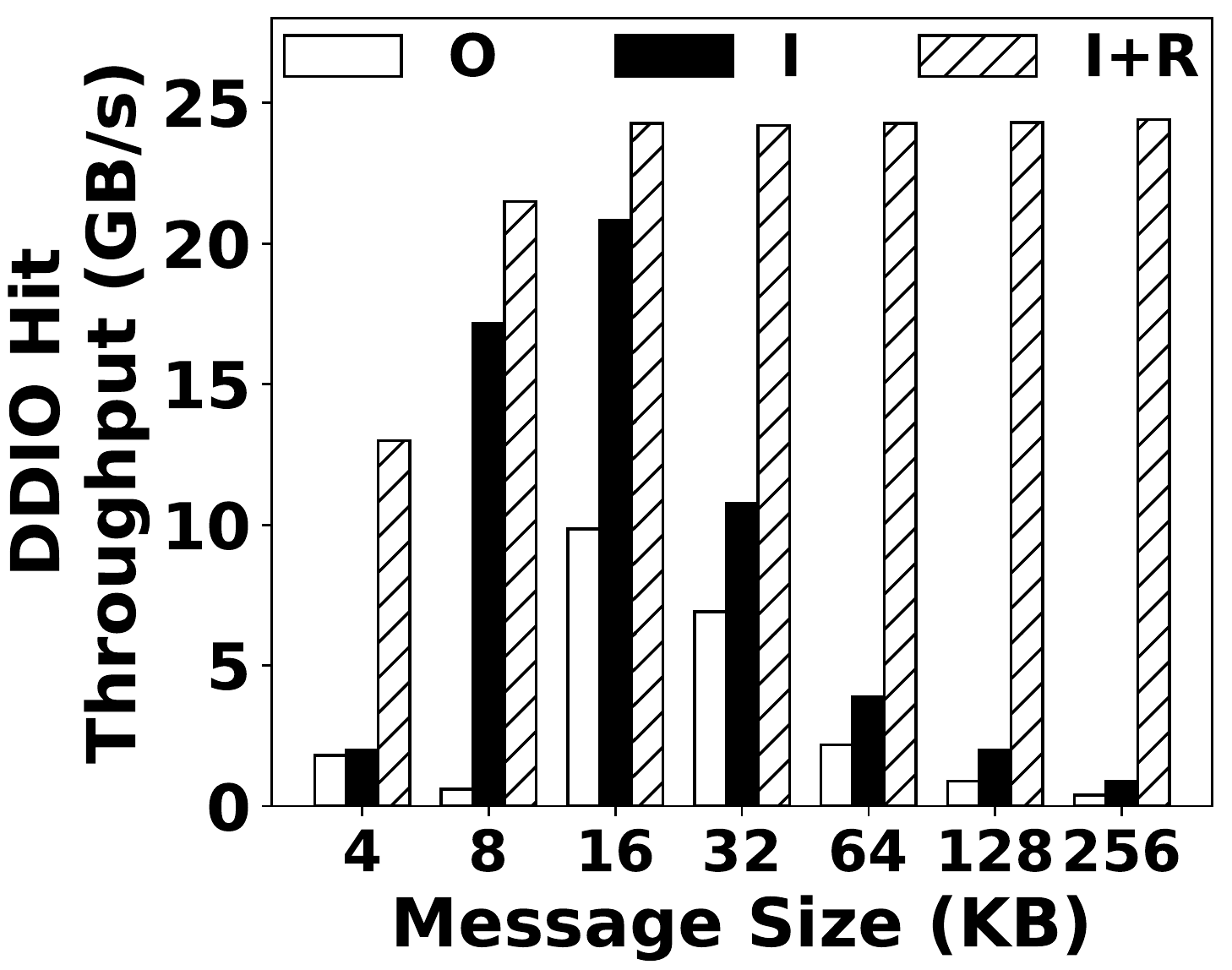}
		\caption{DDIO hit throughput.}
		\label{fig:eval-micro-100-hit-throughput}
	\end{subfigure}
\quad
  	\begin{subfigure}[t]{0.18\linewidth}
		\centering
		\includegraphics[width=\linewidth]{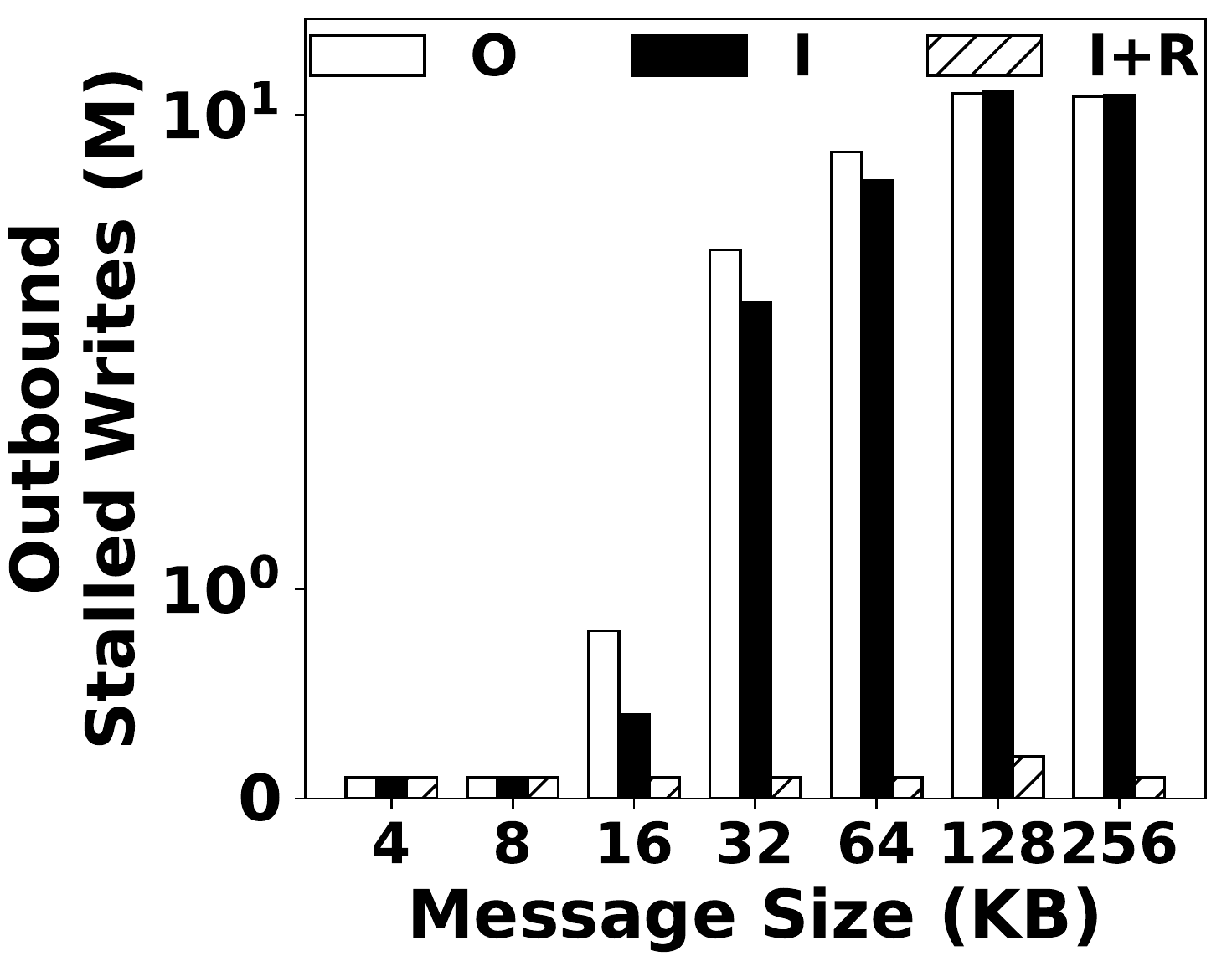}
		\caption{PCIe back pressure.}
		\label{fig:eval-micro-100-pcie}
	\end{subfigure}
	
	\caption{Micro-benchmark experiment results on the dual-port 100~Gbps PFC-free RDMA network.}
	\label{fig:group-100-micro}
\vspace{-2em}
\end{figure*}
The micro-benchmark evaluates and validates the designs in \system{} separately. We first measure the macro metrics to show the overall performance. Under this premise, we explore its inner mechanisms via the micro metrics.

\para{Throughput.} Figure~\ref{fig:eval-micro-25-bandwidth} and~\ref{fig:eval-micro-100-bandwidth} show the results of network throughput results. Overall, the network throughput of  \system{} outperforms the baseline and achieves up to 1.96x throughput. Compared with the baseline, \system{} with cache isolation and cache-resident buffer pool achieves 1.96x and 1.54x throughput in PFC-free networks and PFC-enabled networks with a 256~KB message size, respectively. Compared with the baseline, cache isolation improves network throughput by 5\% and 2.6\%  for PFC-free networks and  networks with PFC on the 256~KB message size. This reveals that \system{} can improve the throughput via bypassing the memory bandwidth bottleneck.

\para{Network latency.} The experimental results of network latency are shown in Figure~\ref{fig:eval-micro-25-latency} and \ref{fig:eval-micro-100-latency}. As we can see, \system{} achieves low latency in both networks, especially in the PFC-enabled network. In the PFC-enabled network, \system{} improves the network delay of small messages~(4~KB) by  46.4\%; the 256~KB messages' delay is improved by 74.8\%. In the PFC-free network, the performance improvement of \system{} is more significant. In these three designs, the cache-resident  buffer pool has the most significant improvement on the latency of the PFC-enabled network and the PFC-free network. The reason is that \system{} can efficiently use cache.

\para{PFC and CNP.} We measure the PFC duration of the PFC-enabled networks and the number of CNP of the PFC-free networks for fine-grained analysis of network throughput. In summary, \system{} with the cache isolation and the cache-resident  buffer pool could almost remove PFC pause frames and CNP packets, and guarantee high throughput ( Figure~\ref{fig:eval-micro-25-pfc} and Figure~\ref{fig:eval-micro-100-cnp}). Specifically, for the PFC-enabled networks,  \system{} can make the network not affected by memory bandwidth bottleneck, and avoid buffer accumulation and PFC generation. Figure~\ref{fig:eval-micro-25-pfc} shows for the PFC-enabled networks, the PFC duration of \system{} is almost 0 for different network messages. In addition, as shown in Figure~\ref{fig:eval-micro-100-cnp}, the number of CNP of  \system{} is also close to 0 for most network messages~(except for 32~KB and 64~KB messages). This indicates that \system{} can bypass memory bandwidth bottlenecks and  guarantee network performance.

\para{DDIO hit throughput.} DDIO hits mean DDIO write update is triggered. DDIO hit throughput under the different network messages as shown in Figure~\ref{fig:eval-micro-25-hit-throughput} and Figure~\ref{fig:eval-micro-100-hit-throughput}. This means the network throughput of DDIO writes an update (aka. cache hits throughput). The improvement of DDIO hit throughput is notable. Compared with the baseline,  \system{} performs 56.6x and 26.7x  DDIO hit throughput for PFC-free networks and the PFC-enabled networks for 256~KB network messages under the memory bandwidth bottleneck. The reason is that as the network message size increases, the required memory footprint size also increases, increasing the probability of cache conflicts. \system{} can achieve a high cache hit ratio and liberate from the memory bandwidth bottleneck.

\para{PCIe back pressure.} 
We study the PCIe back pressure in \system{} in Figure~\ref{fig:eval-micro-25-pcie} and Figure~\ref{fig:eval-micro-100-pcie}. Compared with the baseline, \system{} reduces the PCIe outbound stalled writes by up to 99.1\% and 96.4\% for the PFC-free networks and  the PFC-enabled networks for the 256~KB network message under the memory bandwidth bottleneck, respectively. The reason is that \system{} can process the network messages in the LLC directly and avoid touching the memory. With the efficient data paths, \system{}  is free from memory bandwidth competition.

\subsection{Production Storage Workload in Production DCN}
\label{subsec:eval-small-tesetbed}
\begin{figure}[t]
	\centering
   	\begin{subfigure}[t]{0.46\linewidth}
		\centering
		\includegraphics[width=\linewidth]{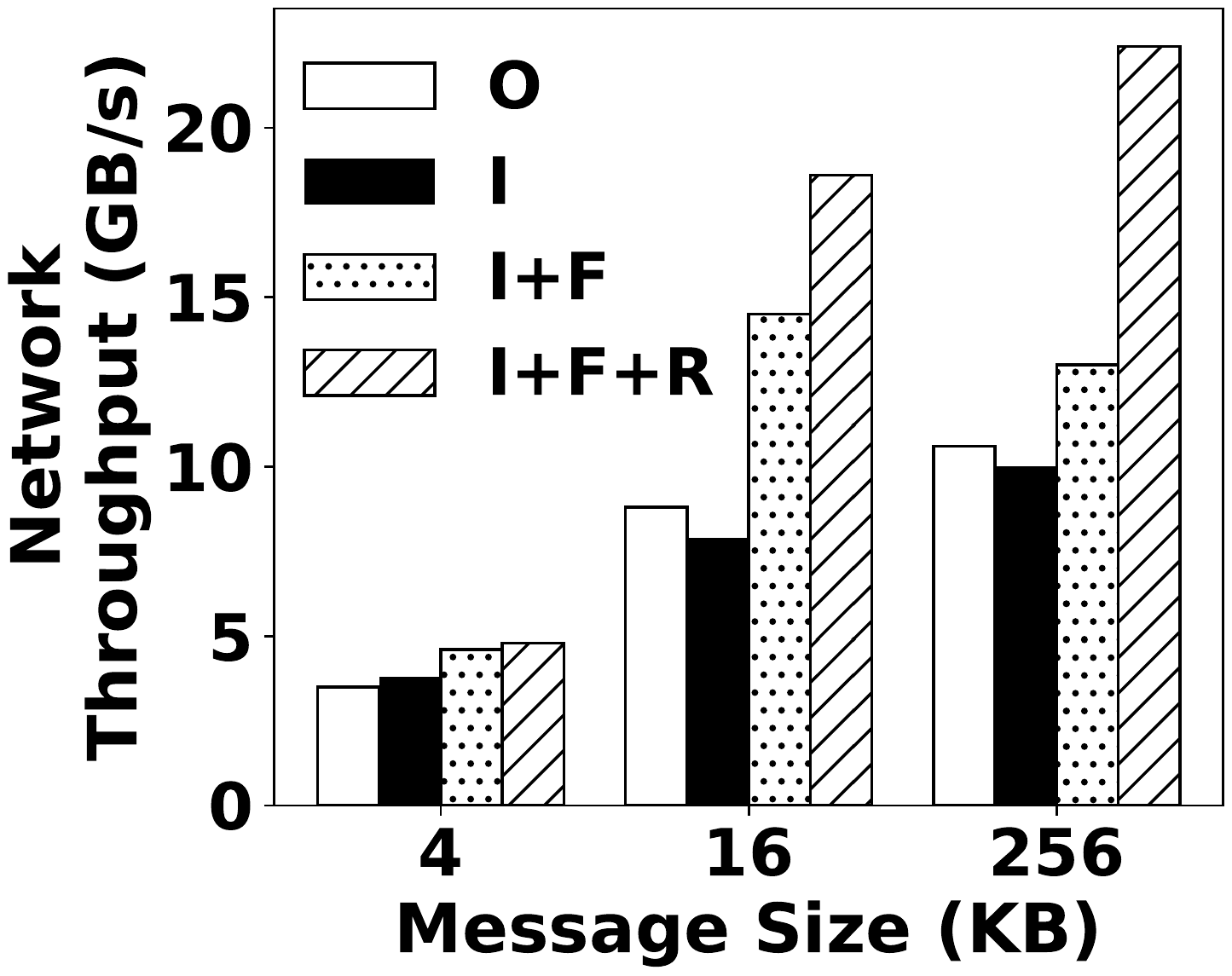}
		\caption{Throughput.}
		\label{fig:eval-macro-net-bandwidth}
	\end{subfigure}
\quad
   	\begin{subfigure}[t]{0.46\linewidth}
		\centering
		\includegraphics[width=\linewidth]{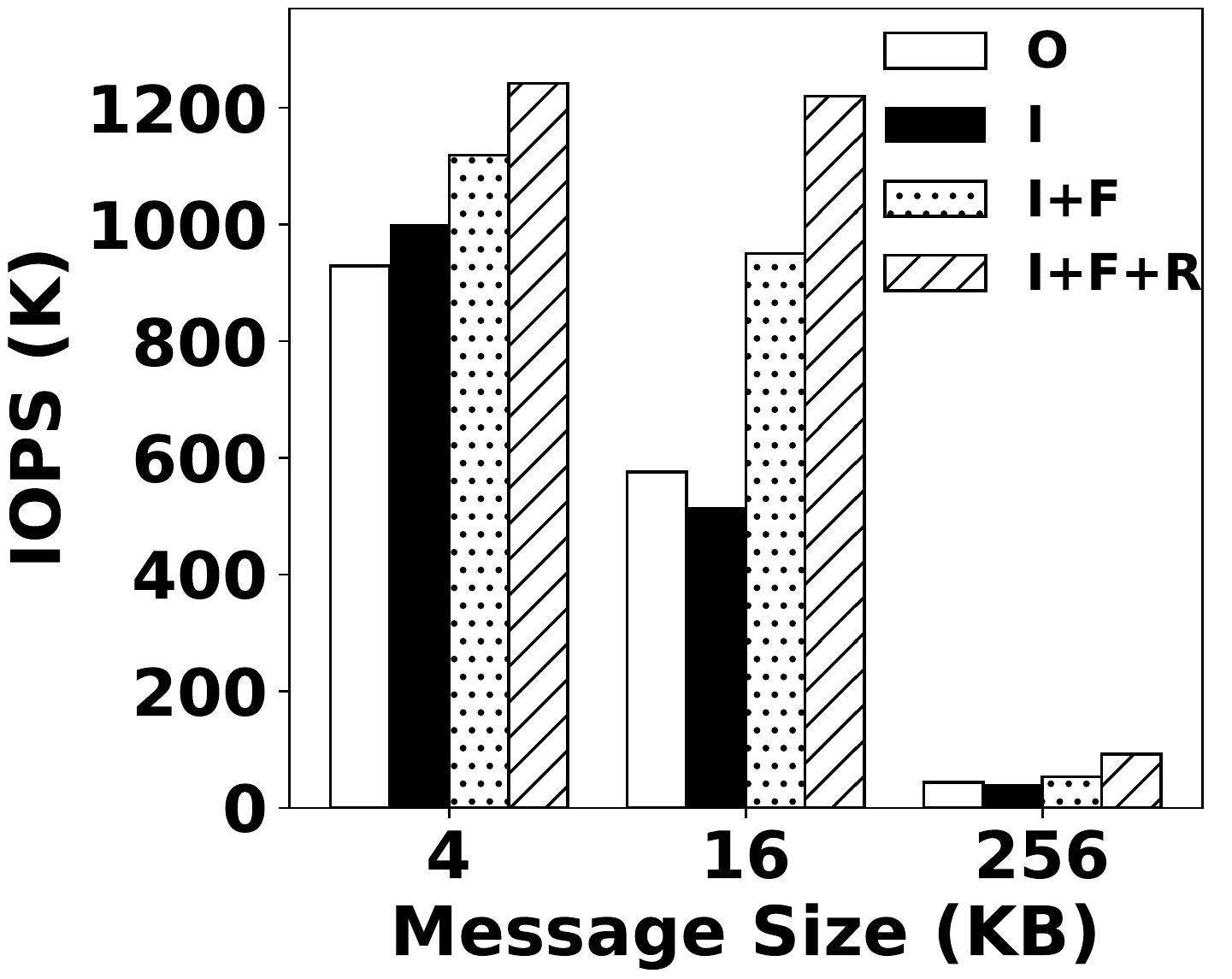}
		\caption{IOPS.}
		\label{fig:eval-macro-iops}
	\end{subfigure}
\quad
   	\begin{subfigure}[t]{0.46\linewidth}
		\centering
		\includegraphics[width=\linewidth]{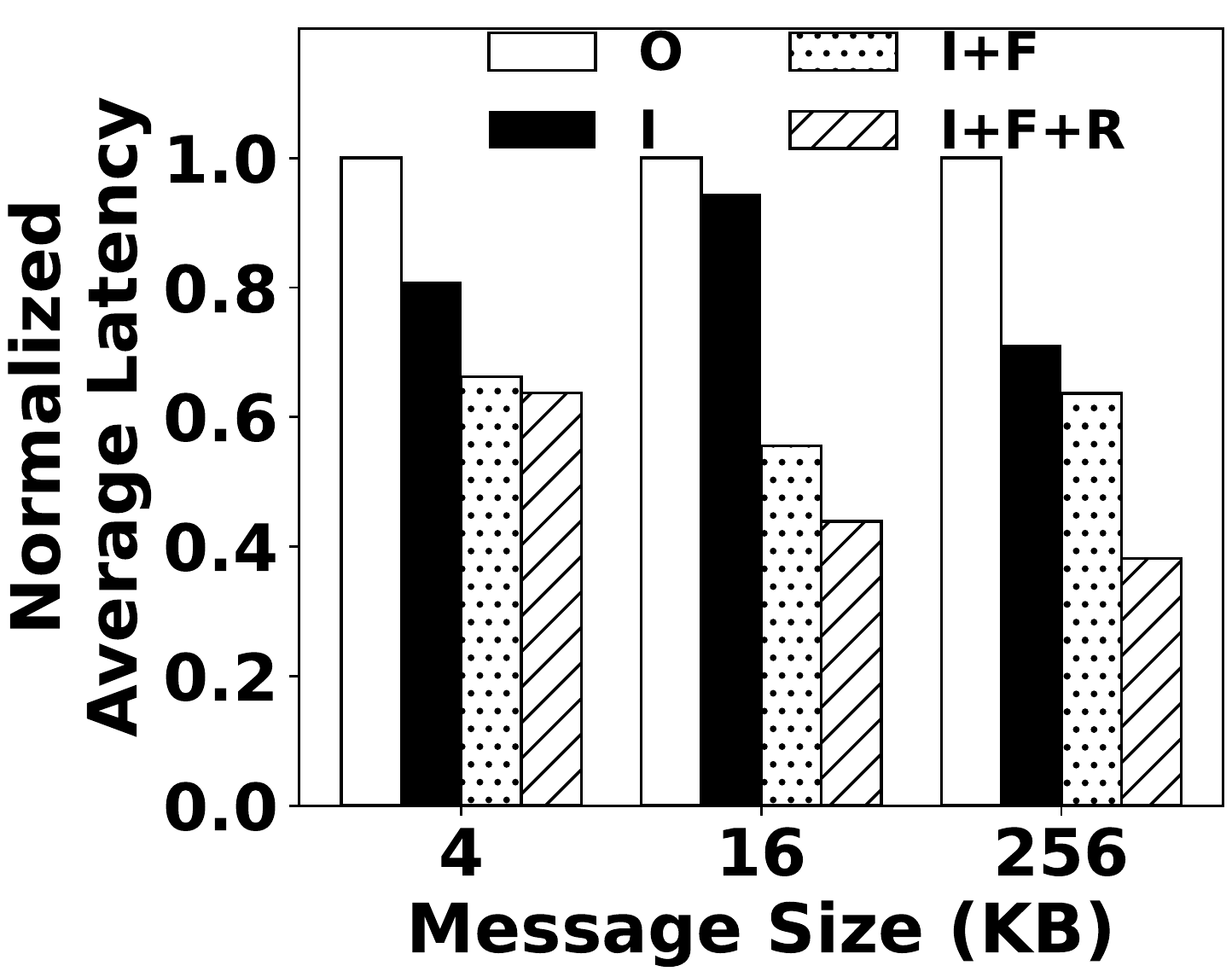}
		\caption{Normalized average latency.}
		\label{fig:eval-macro-latency}
	\end{subfigure}
\quad
   	\begin{subfigure}[t]{0.46\linewidth}
		\centering
		\includegraphics[width=\linewidth]{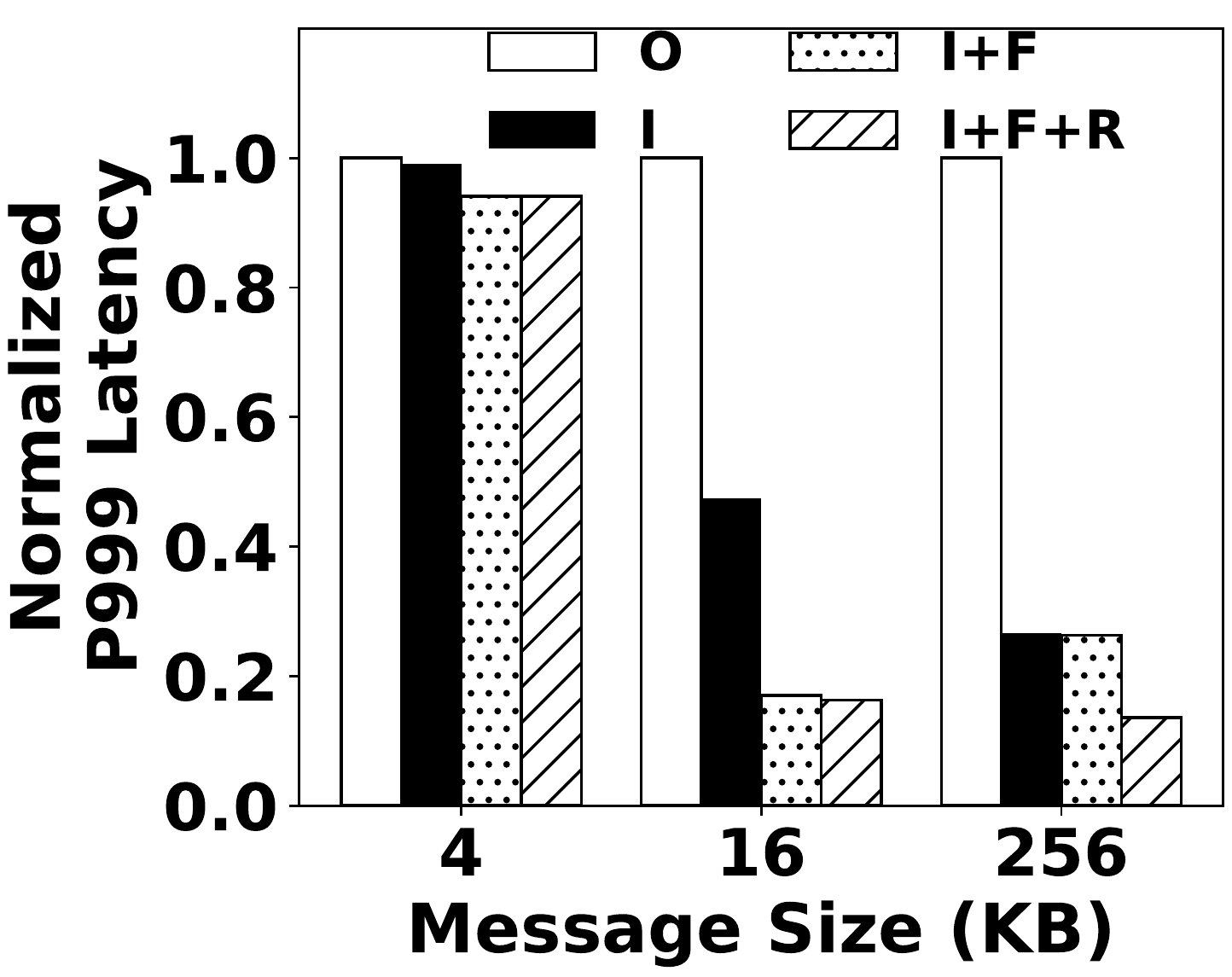}
		\caption{Normalized P999 latency.}
		\label{fig:eval-macro-latency-p999}
	\end{subfigure}
	
	\caption{The performance results of  \system{} and the default design, where `F' means the adaptive cache evacuation.}
	\label{fig:group-macro}
\end{figure}

We evaluate \system{} and the compared methods on multiple macro metrics. Since the three designs of \system{} optimize the performance incrementally, we also evaluate the performance of their incremental combinations in Figure~\ref{fig:group-macro}. We also measure and analyze the running cache and memory of \system{}.

{Throughput.} Figure~\ref{fig:eval-macro-net-bandwidth} shows that compared to the baseline, the CAT cache isolation of \system{} achieves almost the same network throughput with a slight drop for network messages of three sizes. The reason is that the cache isolation reduces the cache used by the network thread, and improves the probability of cache conflict inside the network buffer. The CAT cache isolation and cache evacuation of \system{} achieve better performance compared with the baseline on 4~KB, 16~KB, and 256~KB messages, respectively. With these three designs, \system{} improves the throughput by 2.11x for both 256~KB messages and 16~KB messages. This shows that \system{} can fill the performance degradation gap of isolation and improve the overall throughput.

\para{IOPS.} Figure~\ref{fig:eval-macro-iops} shows that the IOPS of \system{} outperforms the baseline at network messages of three sizes. Compared to the baseline, cache isolation reduces the IOPS for large messages~(256~KB) by 5.7\% and middle messages~(16~KB) by up to 10.4\%. This reason is  the data ping-pong between memory and cache improves latency due to the cache collision. Compared with cache isolation, after integrating fast cache recycle, network messages can be stored in the SSDs at a higher speed, which reduces the probability of cache collisions. The experimental results reveal that the two designs can improve 20.5\% and 64.8\% IOPS compared to the baseline for large and middle messages, respectively. 
By continuing to integrate the cache control, \system{} increases 109.1\% and 111.4\% IOPS for large and middle messages. This shows \system{} can benefit the network messages of storage on IOPS.

\para{Latency.} The experimental results of network delay are shown in Figure~\ref{fig:eval-macro-latency} and Figure~\ref{fig:eval-macro-latency-p999}. We can conclude that the isolation has some influence on the average network latency, whether the size of network messages is 4~KB, 16~KB, or 256~KB. Cache evaluation and cache-resident buffer pool can greatly reduce the average and P999 network latency. For example, with the three designs, the average latency of 16~KB messages decreases by 43.5\% and the average latency of 256~KB messages decreases by 51.6\%. The P999 latency of 256KB messages decreases by 86.4\%. This reveals the effectiveness of the cache control mechanism.

\begin{figure}[t]
	\centering
    	\begin{subfigure}[t]{0.46\linewidth}
		\centering
		\includegraphics[width=\linewidth]{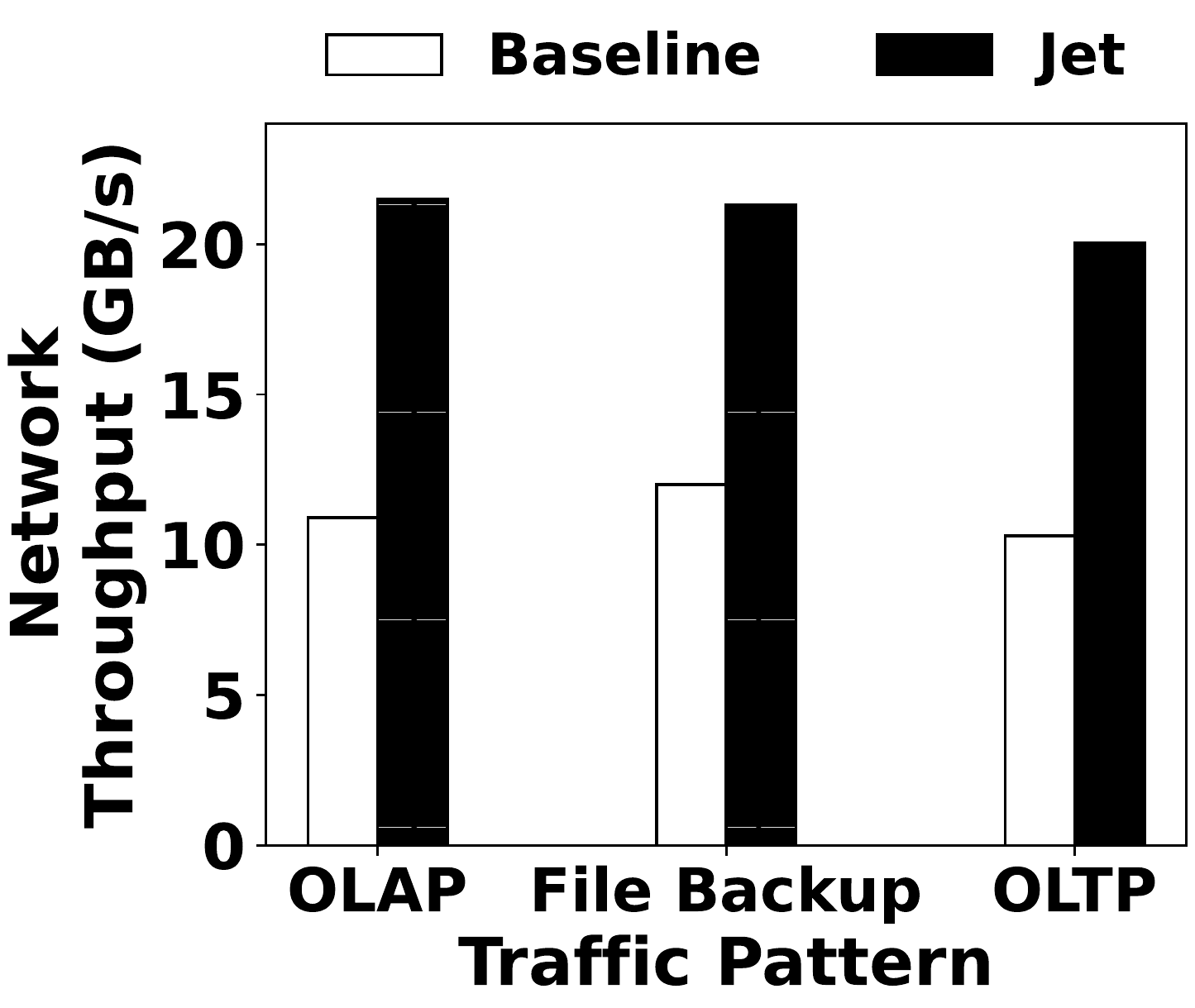}
		\caption{Throughput.}
		\label{fig:eval-real-storage-throughput}
	\end{subfigure}
\quad
    	\begin{subfigure}[t]{0.46\linewidth}
		\centering
		\includegraphics[width=\linewidth]{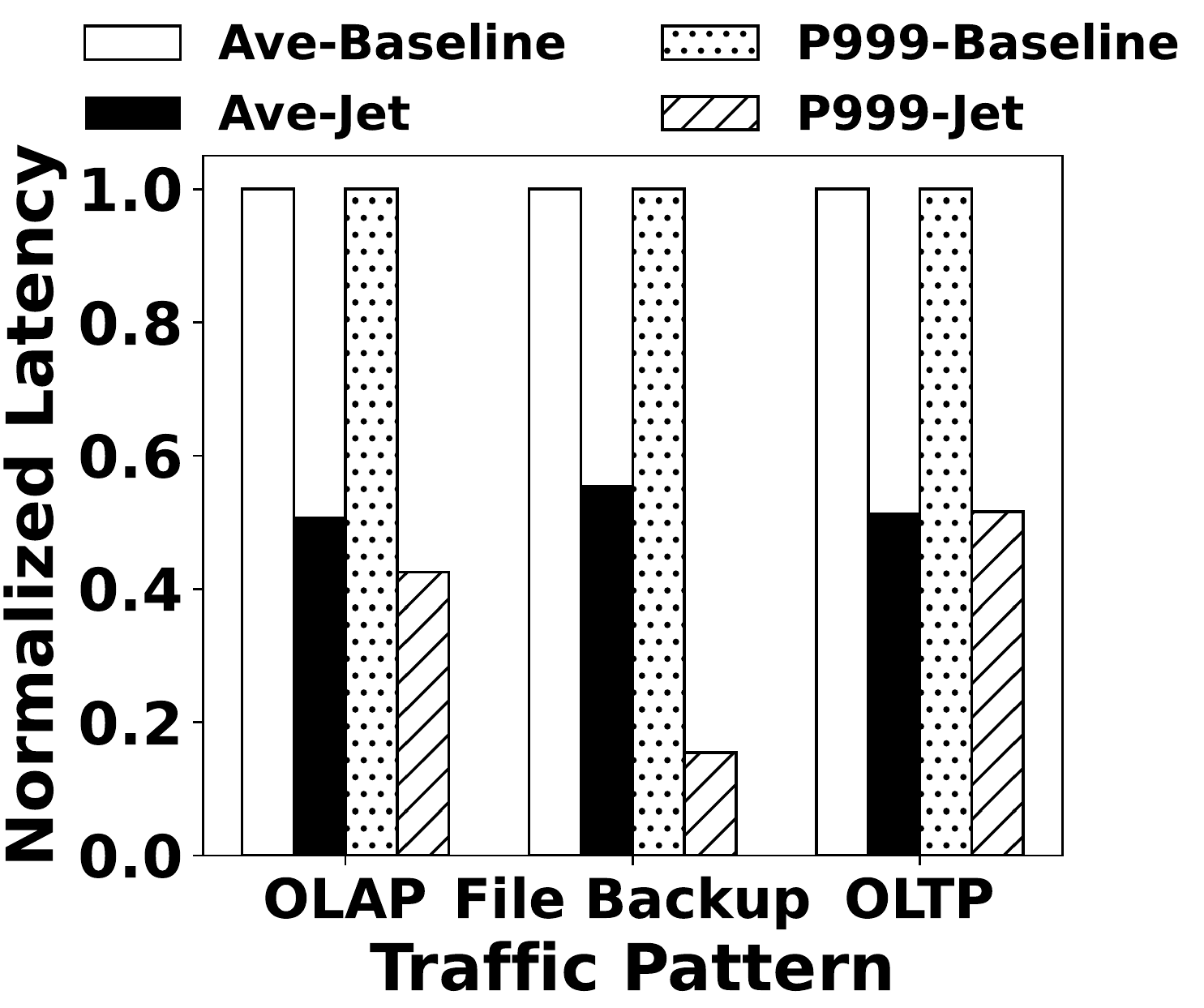}
		\caption{Network latency.}
		\label{fig:eval-real-storage-latency}
	\end{subfigure}
	
	\caption{The \system{} performance in different storage workloads.}
	\label{fig:real-storage}
\end{figure}
\para{The performance of \system{} under different real traffic patterns.} We evaluate the performance of \system{} with different real traffic loads on one of the hosts in the cluster. Specifically, we use three traffic patterns of distributed storage systems: OLAP (online analytical processing), File Backup, and OLTP (online transaction processing), which are monitored and abstracted from the trace of a five-year large-scale cloud storage system. Each traffic pattern is a collection of messages whose size is controlled within a range. We use FIO on one computing node to produce the traffic loads and send them to the storage node. 
Figure~\ref{fig:eval-real-storage-throughput} and Figure~\ref{fig:eval-real-storage-latency} show the throughput and latency results under different traffic patterns.

We can conclude that \system{} improves the throughput by up to 1.97x and reduces the average latency by up to 1.97x on all the traffic patterns. The consistent performance improvement demonstrates the efficacy and efficiency of \system{}. 
\begin{figure}[t]
\vspace{-10pt}
	\centering
     	\begin{subfigure}[t]{\linewidth}
		\centering
		\includegraphics[width=\linewidth]{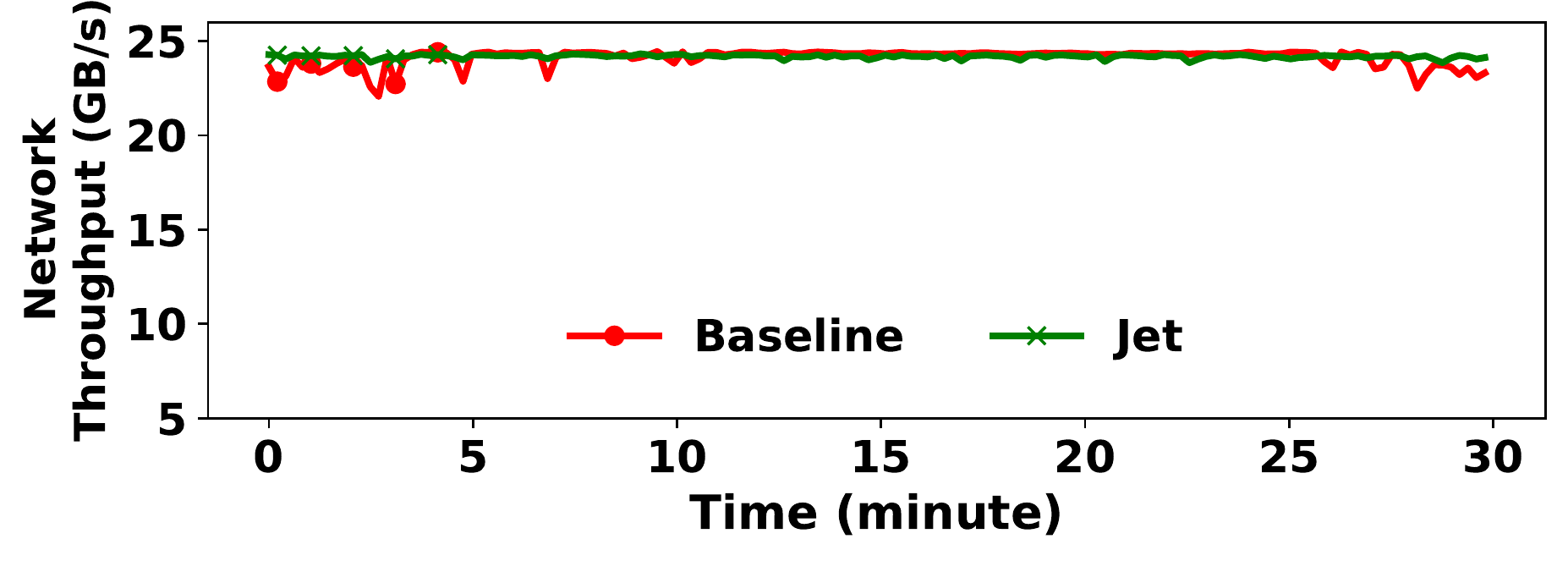}
		\caption{The network throughput comparison for 30 minutes.}
		\label{fig:srq_net_bw}
	\end{subfigure}
      	\begin{subfigure}[t]{\linewidth}
		\centering
		\includegraphics[width=\linewidth]{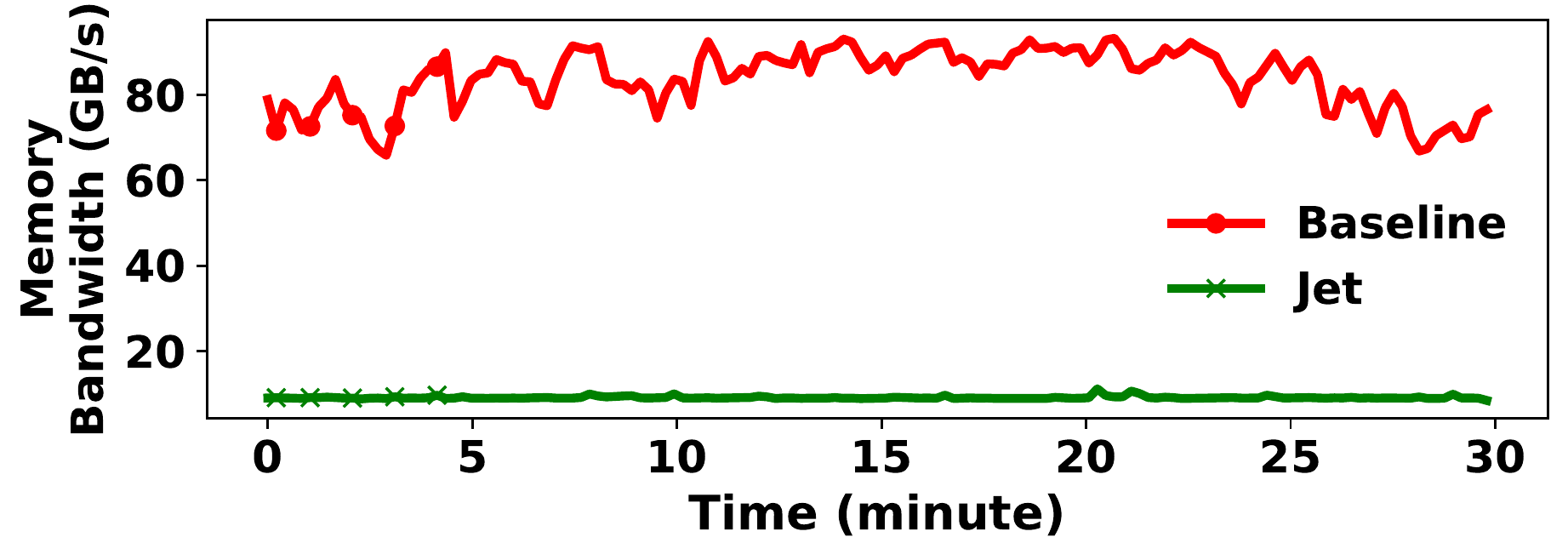}
		\caption{The used memory bandwidth comparison for 30 minutes.}
		\label{fig:srq_mem_bw}
	\end{subfigure}

	\caption{The memory monitor of 
 \sysmethod{} without the memory bandwidth pressure.}
	\label{memory_monitor}
\end{figure}

\begin{figure}[t]
	\centering	
       	\begin{subfigure}[t]{\linewidth}
		\centering
		\includegraphics[width=\linewidth]{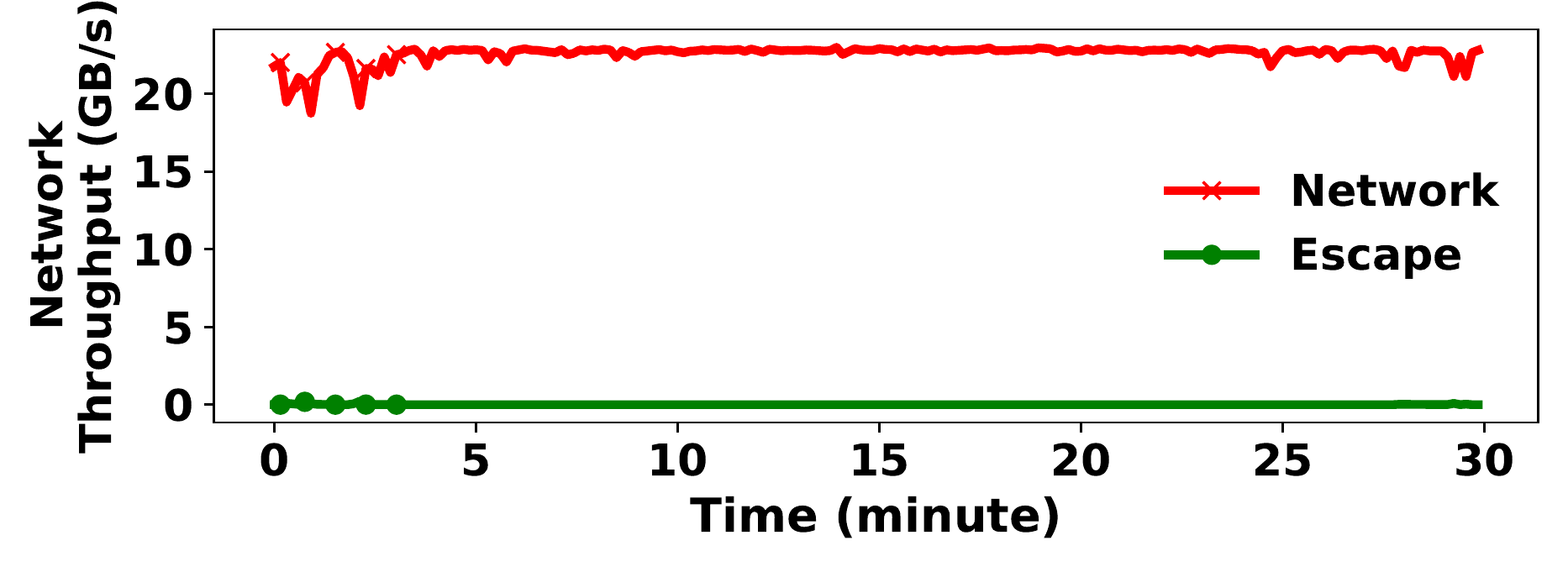}
		\caption{The network throughput of \sysmethod{}.}
		\label{fig:net_throughput}
	\end{subfigure}
        	\begin{subfigure}[t]{\linewidth}
		\centering
		\includegraphics[width=\linewidth]{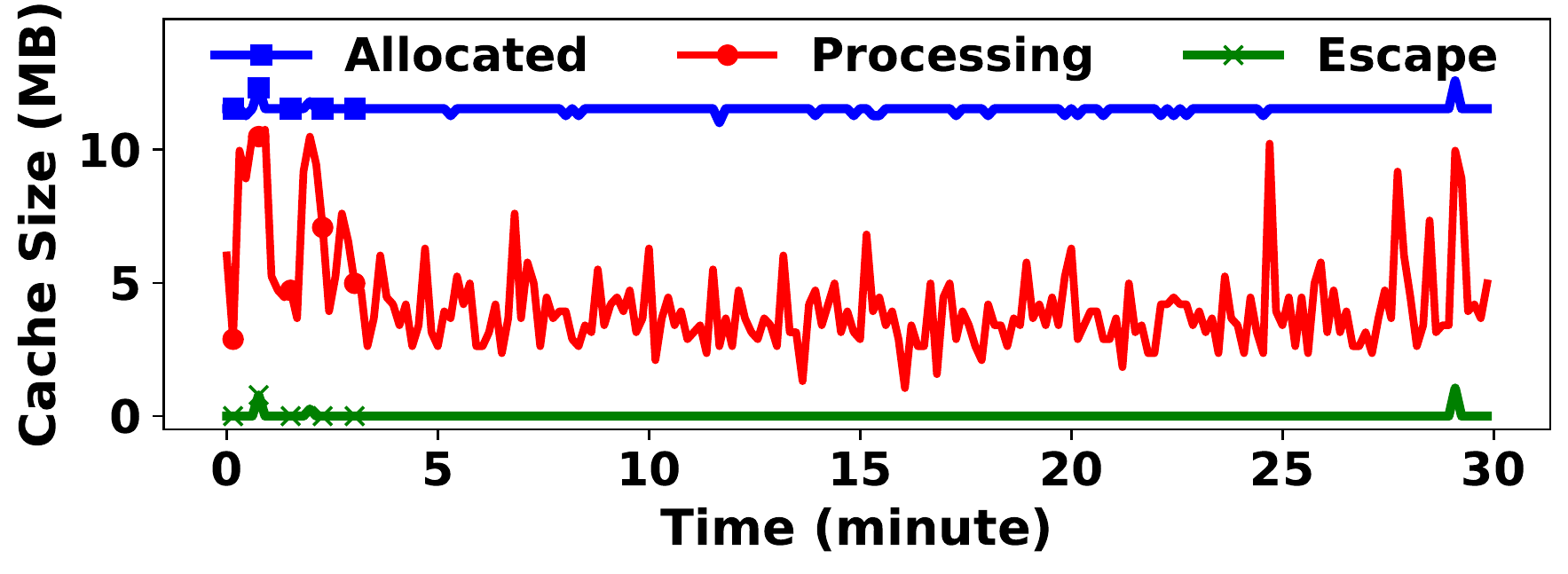}
		\caption{The used cache of \sysmethod{}, where the processing cache, allocated cache, and escape cache indicate the cache that is being occupied by the storage system, the cache pool, and extra cache triggered by the escape mechanics, respectively.}
		\label{fig:escape_cache}
	\end{subfigure}
 
	\caption{The cache monitor of  \sysmethod{}.}
	\label{cache_monitor}
\end{figure}

\para{The used memory bandwidth of  \system{}.} We also measure the memory bandwidth consumption of \system{}. As shown in Figure \ref{fig:srq_net_bw}, \system{} improves network throughput by up to 4.5\% on average. 
We can see that \system{}~decreases nearly 89\% memory bandwidth consumption on average from Figure \ref{fig:srq_mem_bw}. The memory bandwidth consumed by  \system{} is mainly from the escape mechanism. It is worth noting that the memory bandwidth used by the escape mechanism is less than 0.5~GB/s. In summary,  \system{} uses small memory bandwidth to maintain stable network throughput. It reveals the effectiveness of the escape mechanism.
Furthermore,  \system{} can reduce network throughput and memory bandwidth jitters. This is because with \system{}, RNICs can directly write data to the cache and data is forwarded to the SSDs, reducing the number of memory accesses. 

\para{The used cache of \system{}.} We monitor the used cache of \system{} to dig into the internal mechanism of the cache in \system{}. Figure~\ref{fig:escape_cache} shows the cache usage of \system{} without the memory bandwidth pressure. The extra cache triggered by the escape mechanism is less than 1 MB. With such a small escape cache,  \system{} achieves a stable network performance. In Figure~\ref{fig:escape_cache}, the allocated cache and the escape cache are stable except for the start and end. At these two time points,  \system{} triggers the escape mechanism to introduce a cache replacement. 

\begin{figure}[t]
\vspace{-10pt}
	\centering
 	\begin{subfigure}[t]{0.46\linewidth}
		\centering
		\includegraphics[width=\linewidth]{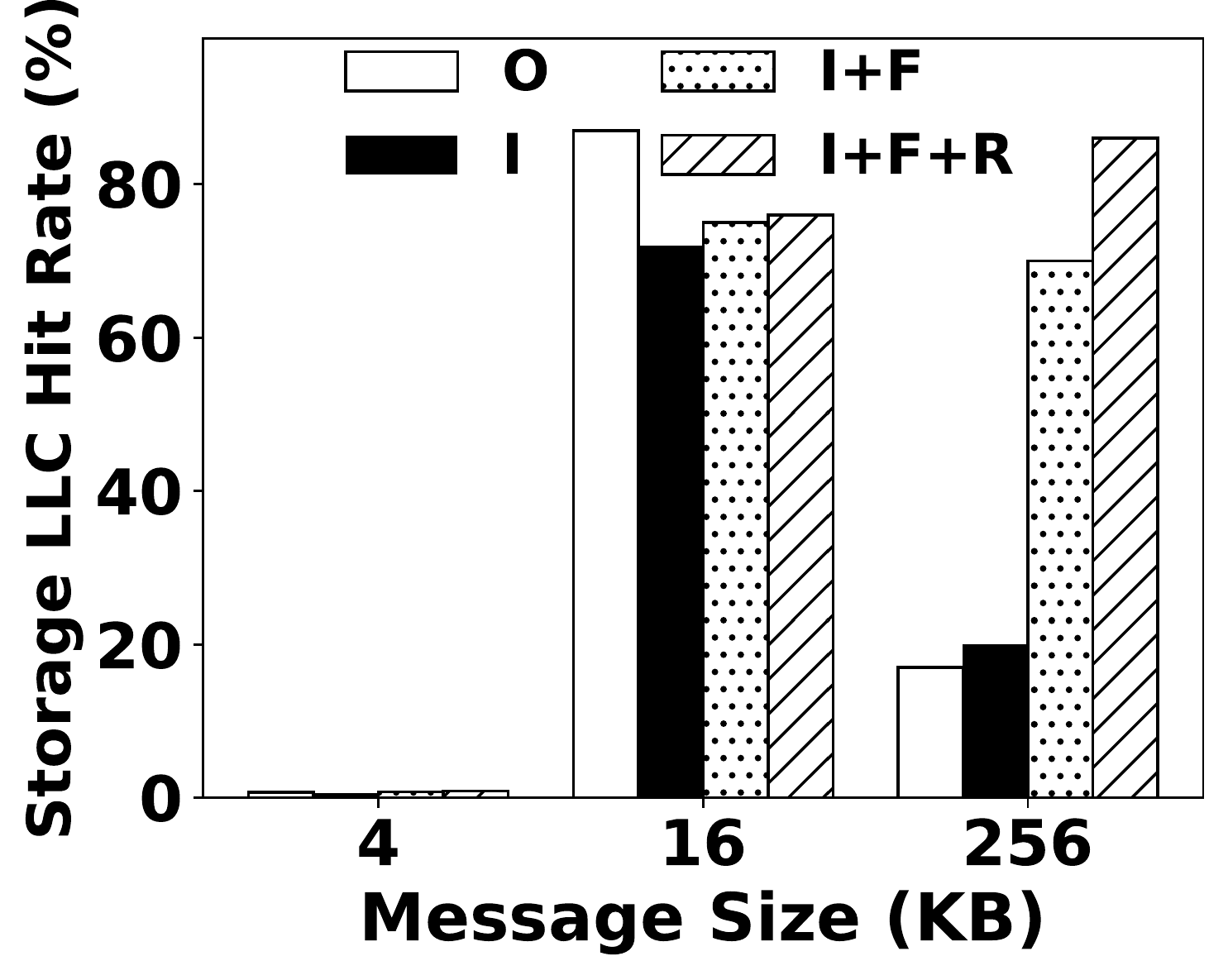}
		\caption{Storage LLC hit ratio.}
		\label{fig:eval-macro-llc-hit-ratio}
	\end{subfigure}
\quad
 	\begin{subfigure}[t]{0.46\linewidth}
		\centering
		\includegraphics[width=\linewidth]{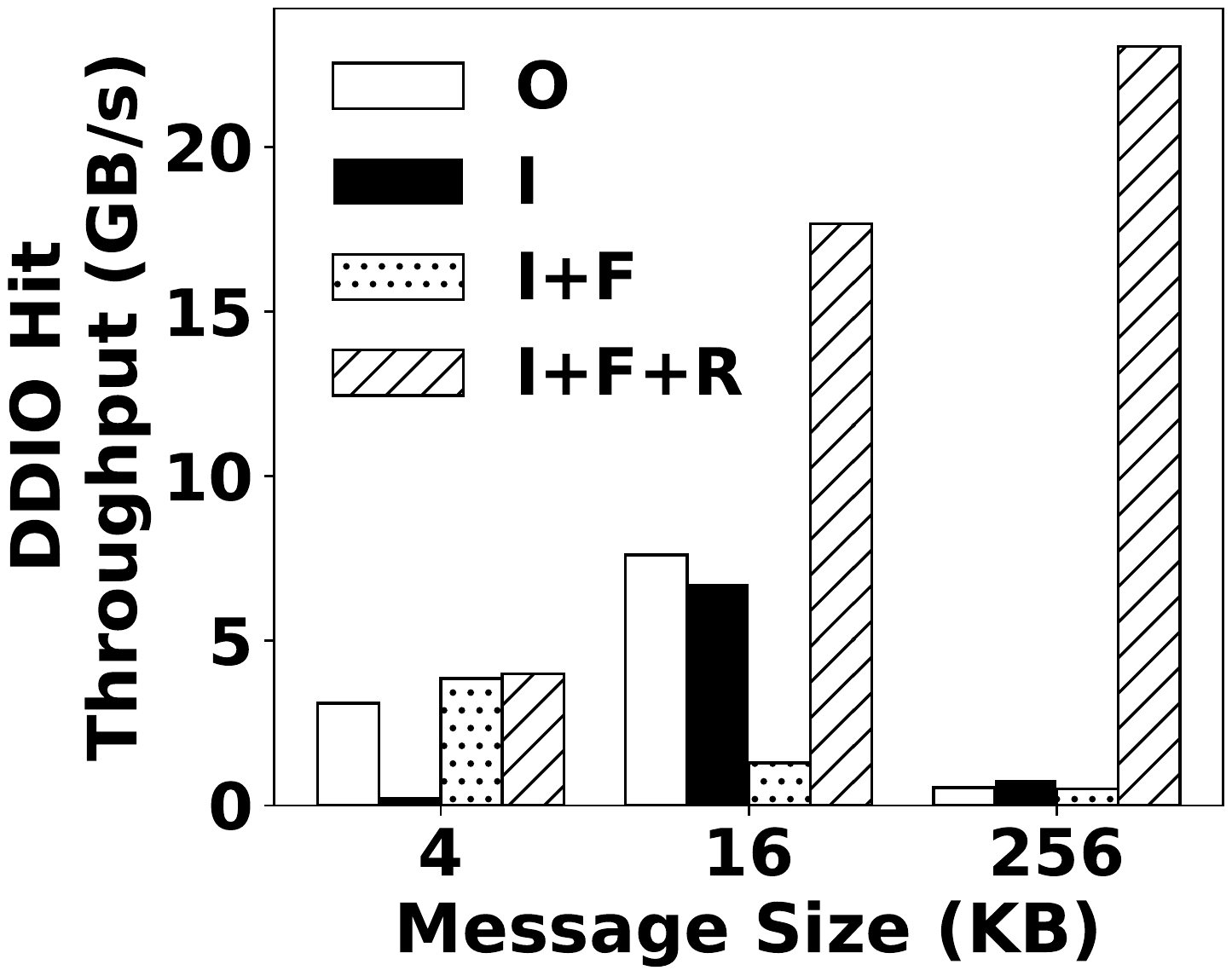}
		\caption{DDIO hit throughput.}
		\label{fig:eval-macro-ddio-hit-ratio}
	\end{subfigure}
 
	\caption{DDIO hit ratio on dual-port 100~Gbps network. }
	\label{fig:group-micro-1}
\end{figure}

\para{Storage LLC hit ratio.} We evaluate the LLC hit ratio to analyze the used cache. Figure~\ref{fig:eval-macro-llc-hit-ratio} shows that \system{} improves the LLC hit ratio. The cache isolation and fast cache recycling of \system{} achieve 4.18x LLC hit ratio for large~(256~KB) messages. \system{} can achieve 5.06x LLC hit ratio for the 256~KB messages. The reason is that the storage system is not sensitive to data-locality access. \system{} decouples network threads and storage threads to reduce cache collision. Compared with the baseline, the cache isolation changes little in the LLC hit ratio as the allocated cache  for storage becomes smaller. 

\para{DDIO hit throughput.} To verify the validity of  \system{}, we also measure the DDIO hit throughput. As shown in Figure~\ref{fig:eval-macro-llc-hit-ratio}, \system{} achieves the best network throughput. Compared with the baseline, \system{} achieves 2.32x and 42.4x higher DDIO hit throughput for 16~KB and 256~KB messages under the memory bandwidth bottleneck, respectively. This shows that the DDIO hit of \system{} in storage applications can perform well as in micro-benchmark.

\subsection{\sysmethod{} in HPC Workload}
\label{subsec:eval-dnn}
\begin{figure}[t]
	\centering
	\includegraphics[width=1\linewidth]{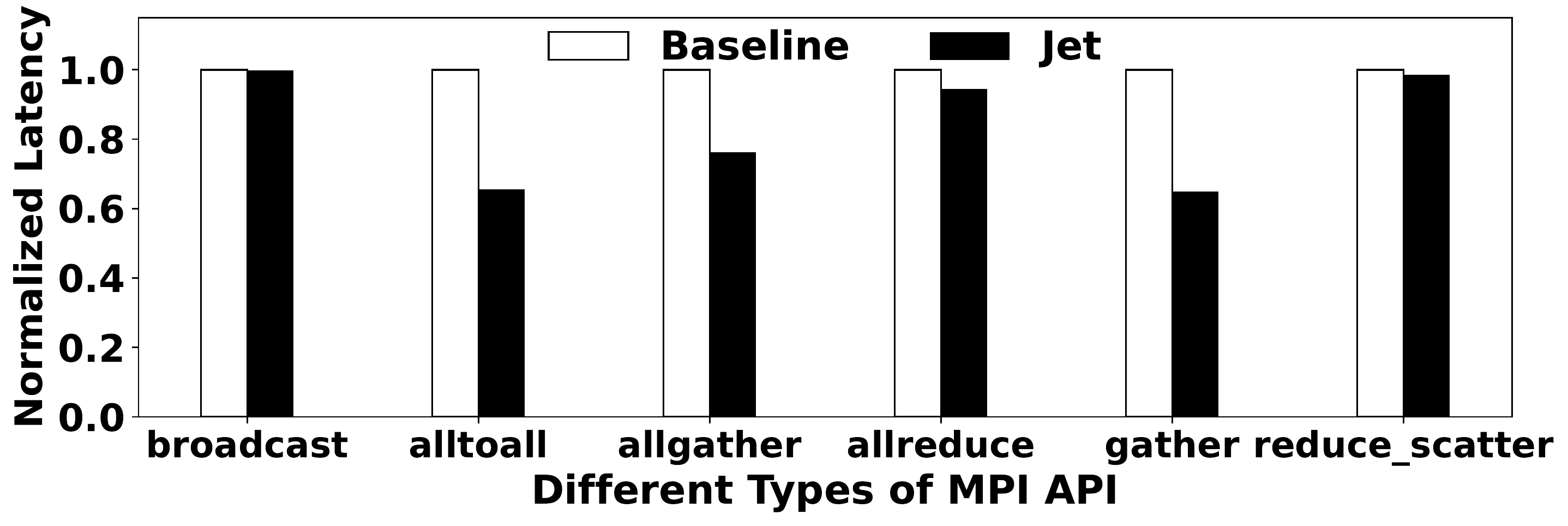}
	\caption{The latency on MPI library benchmark.}
	\label{fig:eval-hpc}
\end{figure}
We also evaluate the performance of  \system{} on  MVAPICH~\cite{mpi}, a high-performance programming library collection. We use two hosts in \S\ref{sec:measurement} to simulate eight computing nodes, where the two hosts are connected with a ToR switch, each host runs four processes, and each process acts as a node. We also configure 12 MB LLC for DDIO and \system{}, where 4 MB accounts for the small messages and 8 MB accounts for the large messages. We run six MPI benchmarks, and each node produces 4 MB network messages. Figure~\ref{fig:eval-hpc} shows the average latency of \system{} compared with  DDIO. \system{}  reduces the communication latency among all nodes under the pressure of the memory bandwidth bottleneck. For the all-to-all and all-gather communication patterns, \system{} reduces the latency by up to 35.1\% and 25\%. For the all-reduce communication pattern, \system{} reduces the latency by up to 5.5\% than the baseline. This reveals that \system{} can improve the performance of HPC applications. 


\section{Related Work}
\label{sec:related}

\para{RDMA operation optimization.} 
Many studies optimize RDMA-enabled distributed applications by using programmable RNICs to offload computation~\cite{DBLP:conf/asplos/KaufmannPSAK16, DBLP:conf/isca/Seemakhupt0SSK21,  DBLP:conf/nsdi/Shu0CGQXCM19, DBLP:conf/eurosys/SidlerWCKA20, aifm, hydra, DBLP:conf/asplos/LazarevXAZD21, hyperloop} and designing scalable RPCs to efficiently use RNIC buffers~\cite{10.1145/3387514.3405897, DBLP:conf/eurosys/ChenLS19, fasst,erpc}. They are tailored for specific applications or specific RDMA verbs. 
DC providers design different congestion
control algorithms~\cite{DBLP:conf/sigcomm/ZhuEFGLLPRYZ15,10.1145/3387514.3405897,acc,kumar2020swift,hpcc,montazeri2018homa} to ensure RDMA operation efficiently in large-scale DCNs. They focus on the network fabric congestion and ignore the potential congestion on the receiver host datapath. 
Some recent studies propose to redesign RNIC by adding memory modules~\cite{netdam, DBLP:conf/micro/AlianK19, DBLP:conf/asplos/PismennyL0T22, DBLP:conf/isca/LimMSRW13}. They require upgrading DCN with specialized, non-commodity RNICs. 
In contrast, we identify that the memory bandwidth bottleneck can cause congestion in the receiver host datapath and propose the RDCA architecture to circumvent it. To our best knowledge,
we are the first to identify and tackle this memory bandwidth bottleneck.

\para{Direct cache access.}
DCA~\cite{DBLP:conf/isca/HuggahalliIT05} and DDIO~\cite{ddio_brief,arm_caceh_stashing} take the first steps in using LLC for fast packet processing. Their benefit in high-speed networks is small due to the leaky DMA problem~\cite{DBLP:conf/micro/VemmouCD22, DBLP:conf/usenix/FarshinRMK20,
DBLP:conf/nsdi/TootoonchianPLW18, DBLP:conf/isca/YuanA00KTK21}.
Some studies propose to dynamically adjust the LLC allocation in response to the cache pressure to improve the performance of DDIO
\cite{DBLP:conf/fpl/FukudaITKSAM14,DBLP:conf/isca/YuanA00KTK21,DBLP:conf/usenix/FarshinRMK20,DBLP:conf/nsdi/TootoonchianPLW18}. In our evaluation, we show that even doubling the LLC size for DDIO does not improve its performance in high-speed networks due to the memory bandwidth bottleneck. In contrast, \system{} fully multiplexes a small area of LLC to achieve efficient RDCA, enabling high throughput and low latency under memory bandwidth bottleneck.

\para{Congestion in the receiver host datapath.} 
Our private conversations with other DC operators show that the memory-bandwidth-induced congestion in the receiver host datapath is not an isolated finding. One operator recently publishes about this issue without proposing a solution
~\cite{DBLP:conf/hotnets/0001AMMERKKRCV22}. In contrast, we propose and implement the RDCA architecture that addresses this issue with success.

\section{Conclusion}\label{sec:conclusion}
We identify the memory bandwidth as the bottleneck of the receiver host datapath
of RDMA. To cope with it, we propose the RDCA architecture that lets remote hosts directly access local cache. We design a
service \system{} that realizes RDCA. Extensive experiments demonstrate its feasibility and benefits.

\para{Acknowledgments.} 
we are extremely grateful for the wonderful comments from the OSDI'23 anonymous reviewers. The comments are invaluable for improving our manuscript in the next version.

\bibliographystyle{unsrt}
\bibliography{sample}

\begin{thebibliography}{10}

\bibitem{DBLP:conf/fast/PanSZSZSPSWGCPS21}
Satadru Pan, Theano Stavrinos, Yunqiao Zhang, Atul Sikaria, Pavel Zakharov,
  Abhinav Sharma, Shiva~Shankar P., Mike Shuey, Richard Wareing, Monika
  Gangapuram, Guanglei Cao, Christian Preseau, Pratap Singh, Kestutis
  Patiejunas, J.~R. Tipton, Ethan Katz{-}Bassett, and Wyatt Lloyd.
\newblock Facebook's tectonic filesystem: Efficiency from exascale.
\newblock In {\em {FAST}}, pages 217--231. {USENIX} Association, 2021.

\bibitem{DBLP:conf/sosp/GhemawatGL03}
Sanjay Ghemawat, Howard Gobioff, and Shun{-}Tak Leung.
\newblock The google file system.
\newblock In {\em {SOSP}}, pages 29--43. {ACM}, 2003.

\bibitem{DBLP:journals/pc/SpiesBMOR22}
Lukas Spies, Amanda Bienz, J.~David Moulton, Luke~N. Olson, and Andrew Reisner.
\newblock Tausch: {A} halo exchange library for large heterogeneous computing
  systems using mpi, opencl, and {CUDA}.
\newblock {\em Parallel Comput.}, 114:102973, 2022.

\bibitem{mpi}
Mvapich.
\newblock \url{https://mvapich.cse.ohio-state.edu/} Accessed Dec 12, 2022.

\bibitem{DBLP:journals/cluster/BawankuleDS22}
Kamalakant~Laxman Bawankule, Rupesh~Kumar Dewang, and Anil~Kumar Singh.
\newblock Historical data based approach to mitigate stragglers from the reduce
  phase of mapreduce in a heterogeneous hadoop cluster.
\newblock {\em Clust. Comput.}, 25(5):3193--3211, 2022.

\bibitem{10.1145/1327452.1327492}
Jeffrey Dean and Sanjay Ghemawat.
\newblock Mapreduce: Simplified data processing on large clusters.
\newblock {\em Commun. ACM}, 51(1):107–113, 2008.

\bibitem{DBLP:conf/osdi/QiaoCSNH0GX21}
Aurick Qiao, Sang~Keun Choe, Suhas~Jayaram Subramanya, Willie Neiswanger,
  Qirong Ho, Hao Zhang, Gregory~R. Ganger, and Eric~P. Xing.
\newblock Pollux: Co-adaptive cluster scheduling for goodput-optimized deep
  learning.
\newblock In {\em {OSDI}}. {USENIX} Association, 2021.

\bibitem{DBLP:conf/osdi/JiangZLYCG20}
Yimin Jiang, Yibo Zhu, Chang Lan, Bairen Yi, Yong Cui, and Chuanxiong Guo.
\newblock A unified architecture for accelerating distributed {DNN} training in
  heterogeneous {GPU/CPU} clusters.
\newblock In {\em {OSDI}}, pages 463--479. {USENIX} Association, 2020.

\bibitem{DBLP:conf/nsdi/GaoLTXZPLWLYFZL21}
Yixiao Gao, Qiang Li, Lingbo Tang, Yongqing Xi, Pengcheng Zhang, Wenwen Peng,
  Bo~Li, Yaohui Wu, Shaozong Liu, Lei Yan, Fei Feng, Yan Zhuang, Fan Liu, Pan
  Liu, Xingkui Liu, Zhongjie Wu, Junping Wu, Zheng Cao, Chen Tian, Jinbo Wu,
  Jiaji Zhu, Haiyong Wang, Dennis Cai, and Jiesheng Wu.
\newblock When cloud storage meets {RDMA}.
\newblock In {\em {NSDI}}, pages 519--533. {USENIX} Association, 2021.

\bibitem{DBLP:conf/sigcomm/GuoWDSYPL16}
Chuanxiong Guo, Haitao Wu, Zhong Deng, Gaurav Soni, Jianxi Ye, Jitu Padhye, and
  Marina Lipshteyn.
\newblock {RDMA} over commodity ethernet at scale.
\newblock In {\em {SIGCOMM}}, pages 202--215. {ACM}, 2016.

\bibitem{DBLP:conf/sigcomm/MittalLDBWGVWWZ15}
Radhika Mittal, Vinh~The Lam, Nandita Dukkipati, Emily~R. Blem, Hassan M.~G.
  Wassel, Monia Ghobadi, Amin Vahdat, Yaogong Wang, David Wetherall, and David
  Zats.
\newblock {TIMELY:} rtt-based congestion control for the datacenter.
\newblock In {\em {SIGCOMM}}, pages 537--550. {ACM}, 2015.

\bibitem{DBLP:conf/nsdi/KongZZJYGZ22}
Xinhao Kong, Yibo Zhu, Huaping Zhou, Zhuo Jiang, Jianxi Ye, Chuanxiong Guo, and
  Danyang Zhuo.
\newblock Collie: Finding performance anomalies in {RDMA} subsystems.
\newblock In {\em {NSDI}}, pages 287--305. {USENIX} Association, 2022.

\bibitem{DBLP:conf/sigcomm/ZhuEFGLLPRYZ15}
Yibo Zhu, Haggai Eran, Daniel Firestone, Chuanxiong Guo, Marina Lipshteyn,
  Yehonatan Liron, Jitendra Padhye, Shachar Raindel, Mohamad~Haj Yahia, and
  Ming Zhang.
\newblock Congestion control for large-scale {RDMA} deployments.
\newblock In {\em {SIGCOMM}}, pages 523--536. {ACM}, 2015.

\bibitem{kumar2020swift}
Gautam Kumar, Nandita Dukkipati, Keon Jang, Hassan~MG Wassel, Xian Wu, Behnam
  Montazeri, Yaogong Wang, Kevin Springborn, Christopher Alfeld, Michael Ryan,
  et~al.
\newblock Swift: Delay is simple and effective for congestion control in the
  datacenter.
\newblock In {\em {SIGCOMM}}, pages 514--528, 2020.

\bibitem{10.1145/3387514.3405897}
Arjun Singhvi, Aditya Akella, Dan Gibson, Thomas~F. Wenisch, Monica Wong-Chan,
  Sean Clark, Milo M.~K. Martin, Moray McLaren, Prashant Chandra, Rob Cauble,
  Hassan M.~G. Wassel, Behnam Montazeri, Simon~L. Sabato, Joel Scherpelz, and
  Amin Vahdat.
\newblock 1rma: Re-envisioning remote memory access for multi-tenant
  datacenters.
\newblock In {\em {SIGCOMM}}, pages 708–--721. Association for Computing
  Machinery, 2020.

\bibitem{DBLP:conf/eurosys/ChenLS19}
Youmin Chen, Youyou Lu, and Jiwu Shu.
\newblock Scalable {RDMA} {RPC} on reliable connection with efficient resource
  sharing.
\newblock In {\em {EuroSys}}, pages 19:1--19:14. {ACM}, 2019.

\bibitem{DBLP:conf/nsdi/Shu0CGQXCM19}
Ran Shu, Peng Cheng, Guo Chen, Zhiyuan Guo, Lei Qu, Yongqiang Xiong, Derek
  Chiou, and Thomas Moscibroda.
\newblock Direct universal access: Making data center resources available to
  {FPGA}.
\newblock In {\em {NSDI}}, pages 127--140. {USENIX} Association, 2019.

\bibitem{DBLP:conf/fpl/FukudaITKSAM14}
Eric~Shun Fukuda, Hiroaki Inoue, Takashi Takenaka, Dahoo Kim, Tsunaki Sadahisa,
  Tetsuya Asai, and Masato Motomura.
\newblock Caching memcached at reconfigurable network interface.
\newblock In {\em {FPL}}, pages 1--6. {IEEE}, 2014.

\bibitem{DBLP:conf/asplos/KaufmannPSAK16}
Antoine Kaufmann, Simon Peter, Naveen~Kr. Sharma, Thomas~E. Anderson, and
  Arvind Krishnamurthy.
\newblock High performance packet processing with flexnic.
\newblock In {\em {ASPLOS}}, pages 67--81. {ACM}, 2016.

\bibitem{DBLP:conf/hotnets/0001AMMERKKRCV22}
Saksham Agarwal, Rachit Agarwal, Behnam Montazeri, Masoud Moshref, Khaled
  Elmeleegy, Luigi Rizzo, Marc~Asher de~Kruijf, Gautam Kumar, Sylvia Ratnasamy,
  David~E. Culler, and Amin Vahdat.
\newblock Understanding host interconnect congestion.
\newblock In {\em {HotNets}}, pages 198--204. {ACM}, 2022.

\bibitem{recio2007remote}
Renato Recio, Bernard Metzler, Paul Culley, Jeff Hilland, and Dave Garcia.
\newblock A remote direct memory access protocol specification.
\newblock Technical report, 2007.

\bibitem{aifm}
Zhenyuan Ruan, Malte Schwarzkopf, Marcos~K. Aguilera, and Adam Belay.
\newblock Aifm: High-performance, application-integrated far memory.
\newblock OSDI'20, 2020.

\bibitem{hydra}
Youngmoon Lee, Hasan~Al Maruf, Mosharaf Chowdhury, Asaf Cidon, and Kang~G.
  Shin.
\newblock Hydra : Resilient and highly available remote memory.
\newblock FAST'22, February 2022.

\bibitem{cx5_manual}
Nvidia connectx-5 ndr 100g infiniband adapter card.
\newblock \url{https://www.nvidia.com/en-us/networking/ethernet/connectx-5/}.

\bibitem{cx7_manual}
Nvidia connectx-7 ndr 400g infiniband adapter card.
\newblock \url{https://nvdam.widen.net/s/srdqzxgdr5/connectx-7-datasheet}.

\bibitem{netdam}
Kevin Fang and David Peng.
\newblock Netdam: Network direct attached memory with programmable in-memory
  computing isa.
\newblock {\em arXiv preprint arXiv:2110.14902}, 2021.

\bibitem{DBLP:conf/micro/AlianK19}
Mohammad Alian and Nam~Sung Kim.
\newblock Netdimm: Low-latency near-memory network interface architecture.
\newblock In {\em {MICRO}}, pages 699--711. {ACM}, 2019.

\bibitem{DBLP:conf/asplos/PismennyL0T22}
Boris Pismenny, Liran Liss, Adam Morrison, and Dan Tsafrir.
\newblock The benefits of general-purpose on-nic memory.
\newblock In {\em {ASPLOS}}, pages 1130--1147. {ACM}, 2022.

\bibitem{little_law}
Arnold~O Allen.
\newblock {\em Probability, statistics, and queueing theory}, page 259.
\newblock Gulf Professional Publishing, 1990.

\bibitem{DBLP:conf/isca/HuggahalliIT05}
Ram Huggahalli, Ravi~R. Iyer, and Scott Tetrick.
\newblock Direct cache access for high bandwidth network {I/O}.
\newblock In {\em {ISCA}}, pages 50--59. {IEEE} Computer Society, 2005.

\bibitem{ddio_brief}
Intel® data direct i/o technology (intel® ddio): A primer.
\newblock
  \url{https://www.intel.com/content/www/us/en/io/data-direct-i-o-technology.html}
  Accessed Nov 30, 2022.

\bibitem{arm_caceh_stashing}
Arm cache stashing.
\newblock
  \url{https://developer.arm.com/documentation/102407/0100/Cache-stashing}
  Accessed Nov 30, 2022.

\bibitem{DBLP:conf/micro/VemmouCD22}
Marina Vemmou, Albert Cho, and Alexandros Daglis.
\newblock Patching up network data leaks with sweeper.
\newblock In {\em {MICRO}}, pages 464--479. {IEEE}, 2022.

\bibitem{DBLP:conf/usenix/FarshinRMK20}
Alireza Farshin, Amir Roozbeh, Gerald Q.~Maguire Jr., and Dejan Kostic.
\newblock Reexamining direct cache access to optimize {I/O} intensive
  applications for multi-hundred-gigabit networks.
\newblock In {\em {ATC}}, pages 673--689. {USENIX} Association, 2020.

\bibitem{DBLP:conf/nsdi/TootoonchianPLW18}
Amin Tootoonchian, Aurojit Panda, Chang Lan, Melvin Walls, Katerina~J.
  Argyraki, Sylvia Ratnasamy, and Scott Shenker.
\newblock Resq: Enabling slos in network function virtualization.
\newblock In {\em {NSDI}}, pages 283--297. {USENIX} Association, 2018.

\bibitem{DBLP:conf/isca/YuanA00KTK21}
Yifan Yuan, Mohammad Alian, Yipeng Wang, Ren Wang, Ilia Kurakin, Charlie Tai,
  and Nam~Sung Kim.
\newblock Don't forget the {I/O} when allocating your {LLC}.
\newblock In {\em {ISCA}}, pages 112--125. {IEEE}, 2021.

\bibitem{amazon-ec2}
Elastic compute cloud (ec2).
\newblock \url{https://aws.amazon.com/ec2} Accessed Dec 12, 2022.

\bibitem{pfc}
IEEE. 802.11Qbb.
\newblock Priority based flow control, 2011.

\bibitem{Intel_rdt}
Intel® resource director technology (intel® rdt), 2021.
\newblock
  \url{https://www.intel.cn/content/www/cn/zh/architecture-and-technology/resource-director-technology.html}
  Accessed Dec 13, 2022.

\bibitem{neo-host}
Mellanox neo-host.
\newblock
  \url{https://support.mellanox.com/s/productdetails/a2v50000000N2OlAAK/mellanox-neohost}
  Accessed Dec 12, 2022.

\bibitem{intel-pcm}
Intel® performance counter monitor.
\newblock \url{https://www.intel.com/software/pcm} Accessed Dec 12, 2022.

\bibitem{intel_e5}
Intel® xeon® processor e5.
\newblock
  \url{https://ark.intel.com/content/www/us/en/ark/products/64597/intel-xeon-processor-e52665-20m-cache-2-40-ghz-8-00-gts-intel-qpi.html}.

\bibitem{adm_cpu}
Amd 3rd gen amd epyc ™ cpus.
\newblock \url{https://www.amd.com/en/claims/epyc3x} Accessed Nov 30, 2022.

\bibitem{ib}
Infiniband trade association. infinibandtm architecture specification volume 1,
  release 1.6, published july 15, 2022.

\bibitem{196243}
Anuj Kalia, Michael Kaminsky, and David~G. Andersen.
\newblock Design guidelines for high performance {RDMA} systems.
\newblock In {\em {ATC}}, pages 437--450. USENIX Association, 2016.

\bibitem{montazeri2018homa}
Behnam Montazeri, Yilong Li, Mohammad Alizadeh, and John Ousterhout.
\newblock Homa: A receiver-driven low-latency transport protocol using network
  priorities.
\newblock In {\em {SIGCOMM}}, pages 221--235, 2018.

\bibitem{252795}
Jeff Bonwick.
\newblock The slab allocator: An {Object-Caching} kernel.
\newblock In {\em {USENIX Summer 1994 Technical Conference}}. USENIX
  Association, 1994.

\bibitem{mellanoxcx-5}
Nvidia® mellanox® connectx®-5.
\newblock \url{https://www.nvidia.com/en-us/networking/ethernet/connectx-5/}
  Accessed Dec 12, 2022.

\bibitem{DBLP:conf/usenix/LiYDMLH20}
Huiba Li, Yifan Yuan, Rui Du, Kai Ma, Lanzheng Liu, and Windsor Hsu.
\newblock {DADI:} block-level image service for agile and elastic application
  deployment.
\newblock In {\em {ATC}}, pages 727--740. {USENIX} Association, 2020.

\bibitem{protobuf}
Google protocol buffers.
\newblock \url{https://developers.google.com/protocol-buffers/} Accessed Dec
  12, 2022.

\bibitem{conf/sigcomm/MiaoZMQZLCGZZLS22}
Rui Miao, Lingjun Zhu, Shu Ma, Kun Qian, Shujun Zhuang, Bo~Li, Shuguang Cheng,
  Jiaqi Gao, Yan Zhuang, Pengcheng Zhang, Rong Liu, Chao Shi, Binzhang Fu,
  Jiaji Zhu, Jiesheng Wu, Dennis Cai, and Hongqiang~Harry Liu.
\newblock From luna to solar: the evolutions of the compute-to-storage networks
  in alibaba cloud.
\newblock In {\em {SIGCOMM}}, pages 753--766. {ACM}, 2022.

\bibitem{connectx-6dx}
Nvidia® connectx®-6 dx.
\newblock \url{https://www.nvidia.com/en-us/networking/ethernet/connectx-6-dx/}
  Accessed Dec 12, 2022.

\bibitem{cat}
Introduction to cache allocation technology in the intel xeon processor e5 v4
  family.
\newblock
  \url{https://www.intel.cn/content/www/cn/zh/developer/articles/technical/introduction-to-cache-allocation-technology.html}
  Accessed Nov 30, 2022.

\bibitem{fio}
Fio.
\newblock \url{https://github.com/axboe/fio} Accessed Nov 22, 2022.

\bibitem{DBLP:conf/isca/Seemakhupt0SSK21}
Korakit Seemakhupt, Sihang Liu, Yasas Senevirathne, Muhammad Shahbaz, and
  Samira~Manabi Khan.
\newblock Pmnet: In-network data persistence.
\newblock In {\em {ISCA}}, pages 804--817. {IEEE}, 2021.

\bibitem{DBLP:conf/eurosys/SidlerWCKA20}
David Sidler, Zeke Wang, Monica Chiosa, Amit Kulkarni, and Gustavo Alonso.
\newblock Strom: smart remote memory.
\newblock In {\em EuroSys}, pages 29:1--29:16. {ACM}, 2020.

\bibitem{DBLP:conf/asplos/LazarevXAZD21}
Nikita Lazarev, Shaojie Xiang, Neil Adit, Zhiru Zhang, and Christina
  Delimitrou.
\newblock Dagger: efficient and fast rpcs in cloud microservices with
  near-memory reconfigurable nics.
\newblock In {\em {ASPLOS}}, pages 36--51. {ACM}, 2021.

\bibitem{hyperloop}
Daehyeok Kim, Amirsaman Memaripour, Anirudh Badam, Yibo Zhu, Hongqiang~Harry
  Liu, Jitu Padhye, Shachar Raindel, Steven Swanson, Vyas Sekar, and Srinivasan
  Seshan.
\newblock Hyperloop: Group-based nic-offloading to accelerate replicated
  transactions in multi-tenant storage systems.
\newblock SIGCOMM'18, 2018.

\bibitem{fasst}
Anuj Kalia, Michael Kaminsky, and David~G. Andersen.
\newblock Fasst: Fast, scalable and simple distributed transactions with
  two-sided (rdma) datagram rpcs.
\newblock OSDI'16, 2016.

\bibitem{erpc}
Anuj Kalia, Michael Kaminsky, and David~G. Andersen.
\newblock Datacenter rpcs can be general and fast.
\newblock NSDI'19, 2019.

\bibitem{acc}
Siyu Yan, Xiaoliang Wang, Xiaolong Zheng, Yinben Xia, Derui Liu, and Weishan
  Deng.
\newblock Acc: Automatic ecn tuning for high-speed datacenter networks.
\newblock In {\em {SIGCOMM}}. Association for Computing Machinery, 2021.

\bibitem{hpcc}
Yuliang Li, Rui Miao, Hongqiang~Harry Liu, Yan Zhuang, Fei Feng, Lingbo Tang,
  Zheng Cao, Ming Zhang, Frank Kelly, Mohammad Alizadeh, et~al.
\newblock Hpcc: High precision congestion control.
\newblock In {\em {SIGCOMM}}, pages 44--58. 2019.

\bibitem{DBLP:conf/isca/LimMSRW13}
Kevin~T. Lim, David Meisner, Ali~G. Saidi, Parthasarathy Ranganathan, and
  Thomas~F. Wenisch.
\newblock Thin servers with smart pipes: designing soc accelerators for
  memcached.
\newblock In {\em {ISCA}}, pages 36--47. {ACM}, 2013.

\end{thebibliography}

\end{document}